%% file: LongPaper.tex
\documentclass[12pt]{article}

\usepackage{amsmath,amsthm,amssymb,stmaryrd}
\usepackage{xspace}
\usepackage{subcaption}
\usepackage[center]{qtree}
\usepackage{natbib}
\usepackage[margin = 3cm]{geometry}
\usepackage{setspace}
\usepackage{hyperref}

\input{tikzpreamble}
\newcommand{\convexrel}{\ensuremath{{\bf ConvexRel}}\xspace}

\newcommand{\freepreg}[1]{\mathsf{Preg}_{#1}}
\newcommand{\lang}[1]{\ensuremath{\textit{#1}}}
\newcommand{\FHilb}{\mathbf{FHilb}}

\newcommand{\vect}[1]{\overrightarrow{#1}}
\newcommand{\ch}[1]{Conv({#1})}
\newcommand{\ket}[1]{\ensuremath{| #1 \rangle}}
\newcommand{\define}[1]{{\bf #1}}
\newcommand{\property}[1]{\ensuremath{\lang{p}_\lang{#1}}}

\newtheorem{thm}{Theorem}

\newtheorem{dfn}{Definition}

\newtheorem{remark}{Remark}
\theoremstyle{definition}\newtheorem{example}{Example}

\let\phi\varphi

\title{Interacting Conceptual Spaces I :\\
  {\large Grammatical Composition of Concepts \footnote{This paper is a significantly extended version of the workshop paper~\cite{BoltCoeckeGenoveseLewisMarsdenPiedeleu2016}}} }
\author{Joe Bolt \quad Bob Coecke\quad Fabrizio Genovese\quad Martha Lewis\\ Dan Marsden \quad Robin Piedeleu \footnote{authors in alphabetical order}\\Quantum Group, Department of Computer Science, University of Oxford}
\date{}

\begin{document}
\maketitle
\begin{abstract}
  The categorical compositional approach to meaning has been successfully applied in natural language processing,
  outperforming other models in mainstream empirical language processing tasks.
  We show how this approach can be generalized to conceptual space models of cognition.
  In order to do this, first  we introduce the category of convex relations as a new setting for categorical compositional semantics,
  emphasizing the convex structure important to conceptual space applications.
  We then show how to construct conceptual spaces for various types such as nouns, adjectives and verbs.
  Finally we show by means of examples how concepts can be systematically combined
  to establish the meanings of composite phrases from the meanings of their constituent parts.   This provides 
  the mathematical underpinnings of a new compositional approach to cognition.
\end{abstract}


\section{Introduction}
  How should we represent concepts and how can they be composed to form new concepts, phrases and sentences? These questions are fundamental to cognitive science and thereby human-like artificial intelligence.
  \em Conceptual spaces theory \em gives a way of describing structured concepts~\citep{Gardenfors2004, Gardenfors2014},
  not starting from linguistic assumptions, but from cognitive considerations about human reasoning.
  Conceptual spaces describe a ``semantics of the mind'', modelling mental descriptions of concepts. 
  The key idea is that human beings represent concepts geometrically in certain fundamental domains of understanding such as space, motion, taste and colour.
  These domains are combined to form a~\emph{conceptual space} describing the features of interest.
  A concept is then described by convex subsets of the relevant domains.
  The convexity requirement can be seen as a means of identifying robust, meaningful concepts. For example, if two points in some space are considered to represent the colour red,
  then intuitively we would expect that every point ``in between'' would also be considered red.
  Conceptual spaces provide a middle ground between symbolic and connectionist representations of concepts, oriented towards tasks of interest in cognitive science,
  such as categorization and assessing similarity.
  
  \em Categorical compositional distributional models \em \citep{CoeckeSadrzadehClark2010} successfully exploit the
  compositional structure of natural language in a principled manner, and have outperformed other approaches in Natural Language Processing (NLP) \citep{GrefSadr, KartSadr}.  The approach works as follows.  
  A mathematical formalization of grammar is chosen, for example \em Lambek's pregroup grammars \em \citep{Lambek1999},
  although the approach is equally effective with other categorial grammars~\citep{CoeckeGrefenstetteSadrzadeh2013}.
  Such a categorial grammar allows one to verify whether a phrase or a sentence is grammatically well-formed by means of a computation that establishes the overall grammatical type,
  referred to as \em a type reduction\em.
  The meanings of \em individual \em words are established using a distributional model of language, where they are described as vectors of co-occurrence statistics derived automatically from corpus data~\citep{Lund1996}.
  The categorical compositional distributional programme unifies these two aspects of language in a compositional model where grammar mediates composition of meanings. This
  allows us to derive the meaning of sentences from their grammatical structure, and the meanings of their constituent words.
  The key insight that allows this approach to succeed is that both pregroups and the category of vector spaces carry the same abstract structure~\citep{CoeckeSadrzadehClark2010},
  and the same holds for other categorial grammars since they typically have a weaker categorical structure.
  

The abstract framework of the categorical compositional scheme we discuss here is broader in scope than just NLP.
It can be applied in other settings in which we wish to compose meanings in a principled manner, guided by structure. 
The outline of the general programme is as follows:
\begin{enumerate}  
  \label{item:compstruct}
\item \begin{enumerate}
\item Choose a compositional structure, such as a pregroup or any other categorial grammar.
\item Interpret this structure as a category, the \define{grammar category}.
\end{enumerate}
\item \begin{enumerate}
  \label{item:meaningspace}
\item Choose or craft appropriate meaning or concept spaces, such as spaces of propositions, vector spaces, density matrices~\citep{Piedeleu2015, Bankova2016} or conceptual spaces.
  \label{item:meaningcategory}
\item Organize these spaces into a category, the \define{semantics category}, with the same abstract structure as the grammar category.
\end{enumerate}
  \label{item:interpret}
\item Interpret the compositional structure of the grammar category in the semantics category via a functor preserving the type reduction
  structure.
  \label{item:reduction}
\item Bingo! This functor maps type reductions in the grammar category onto algorithms for composing meanings in the semantics category. 
\end{enumerate}
In order to move away from vector spaces, we construct a new categorical setting for interpreting meanings which respects the important convex structure emphasized in  conceptual spaces theory.
We show that this category has the necessary abstract structure required by categorical
compositional models. We then construct convex spaces for interpreting the types for nouns, adjective and verbs.
Finally, this allows us to use the reductions of the pregroup grammar to compose meanings in conceptual spaces. We illustrate our approach with concrete examples, and go on to discuss
directions for further research.

\section{Categorical compositional meaning}
In this section we describe the details of the categorical compositional approach to meaning. We provide examples of semantic categories and grammar categories,
and show the general method by which the grammar category induces a notion of concept composition in the semantic category. 

\subsection{Monoidal categories}
We begin by briefly reviewing some of the category theory underpinning categorical compositional models.

\begin{dfn}
A \define{monoidal category} is a tuple $(C, \otimes, I, \alpha, \lambda, \rho)$ where
\begin{itemize}
\item $C$ is a category
\item $\otimes$, the \define{tensor}, is a functor $C\times C\rightarrow C$ where we write $A\otimes B$ for $\otimes (A,B)$
\item $I$, the \define{unit}, is an object of $C$
\end{itemize}
The remaining data are natural isomorphisms, with components of type:
\begin{itemize}
\item $\alpha_{A,B,C} : ((A\otimes B) \otimes C) \rightarrow (A\otimes( B\otimes C))$
\item $\rho_A : A\otimes I \rightarrow A$
\item $\lambda_A : I\otimes A \rightarrow A$
\end{itemize}
These natural isomorphisms, moreover, must be such that any formal and well-typed diagram made up from $\otimes, \alpha, \lambda, \rho, \alpha^{-1}, \rho^{-1}, \lambda^{-1}$ and identities commutes. Here \lq formal\rq \:  means \lq not dependent on the structure of any particular monoidal category.\rq
\end{dfn}

For a precise statement and discussion of the above definition, we direct the reader to~\cite{MacLane1971}. A more gentle introduction can be found in~\cite{CoeckePaquette2011}.  For the purpose of our paper,  the objects of a monoidal category should be thought of as \textit{system types}. A morphism $f:A\rightarrow B$ is then a process taking inputs of type $A$ and giving outputs of type $B$. The object $A\otimes B$ represents the systems~$A$ and~$B$ composed in \textit{parallel}. Hence, a morphism $f\otimes g : A\otimes B\rightarrow C\otimes D$ is to be thought of as running the process $f:A\rightarrow C$ \emph{whilst} running the process $g:B\rightarrow D$. The object $I$ is thought of as the trivial system. 

\begin{example}
The category $\mathbf{Rel}$ of sets and relations is monoidal. The tensor $\otimes$ is the Cartesian product and $I$ is any singleton set $\{ \star \}$. 
\end{example}

\begin{example}
The category $\mathbf{FdVect}_\mathbb{R}$ of finite dimensional real vector  spaces and linear maps is monoidal. The tensor $\otimes$ is the tensor product, the trivial system $I$ is the one-dimensional real vector space $\mathbb{R}$.
\end{example}

Monoidal categories admit an elegant and powerful graphical notation that we will exploit heavily in later sections. In this notation, an object~$A$ is denoted by a wire:
\begin{center}
\input{wire.tikz} 
\end{center}
A morphism $f:A\rightarrow B$ is represented by a box:
\begin{center}
\input{morphism1.tikz}
\end{center}
If $g:B\rightarrow C$, the composite $g\circ f : A\rightarrow C$ is given by:  
\begin{center}
\input{morphism2.tikz}
\end{center}
If $h:A\rightarrow B$ and $k:C\rightarrow D$, the morphism $h\otimes k : A\otimes C \rightarrow B\otimes D$ is depicted by:
\begin{center}
\input{parallel.tikz}
\end{center}
The trivial system $I$ is the empty diagram. Morphisms $u:I\rightarrow A$ and $v: A\rightarrow I$ are drawn respectively as
\begin{equation*}
  \begin{gathered}
    \input{effect.tikz}
  \end{gathered}
  \quad\text{and}\quad
  \begin{gathered}
    \input{effect_copy.tikz}
  \end{gathered}
\end{equation*}
These special morphisms are referred to as~\define{states} and~\define{effects}.
\subsection{Compact closed categories}
A specific class of monoidal categories, the compact closed categories, will be of particular importance.
\begin{dfn}
A monoidal category $(C,\otimes, I)$ is \define{compact closed} if for each object $A\in C$ there are objects $A^l,A^r\in C$ (the \define{left} and \define{right duals} of $A$) and morphisms
\begin{align*}
\eta_A^l : I\rightarrow A\otimes A^l \:&\:& \eta_A^r : I\rightarrow A^r\otimes A
\\
\epsilon_A^l : A^l\otimes A \rightarrow I \:&\:& \epsilon_A^r : A\otimes A^r\rightarrow I
\end{align*}
satisfying the snake equations
\begin{align*}
&(1_A \otimes \epsilon^l )\circ (\eta^l \otimes 1_A) = 1_A &\:& (\epsilon^r \otimes 1_A) \circ (1_A \otimes \eta^r ) = 1_A
\\
&(\epsilon^l \otimes 1_{A^l})\circ (1_{A^l} \otimes \eta^l)= 1_{A^l} &\:& (1_{A^r} \otimes \epsilon^r ) \circ (\eta^r \otimes 1_{A^r} ) = 1_{A^r}
\end{align*}
\end{dfn}
The dual objects are also represented as wires. To distinguish objects and their duals, wires are \emph{directed}. The wire representing an object $A$ is directed down the page, and wires representing dual objects $A^r$ and $A^l$ are directed up the page.

The $\epsilon$ and $\eta$  maps are called \define{caps} and \define{cups} respectively,  and are depicted graphically as:
\begin{center}
\input{cupscaps.tikz}
\end{center}
Graphically, the snake equations become
\begin{center}
\input{snaaaakes.tikz}
\end{center}
Compact closed categories are a convenient level of abstraction at which to work. Many of the categories one would think to use as either grammar or meaning categories have a compact closed structure.
\begin{example}
All objects in \textbf{Rel} are self-dual. Both caps are given by
\begin{gather*}
\epsilon_X : X\otimes X\rightarrow \{\star\} :: \{ \left(( x, x), \star\right) \mid x \in X \} 
\end{gather*}
The associated cup $\eta_X$ is  the converse of the above. The snake equations can be verified by direct calculation.
\end{example}
\begin{example}

$\FHilb$ is the category of finite dimensional real inner product spaces. As in the case of \textbf{FdVect$_\mathbb{R}$}, the tensor $\otimes$ is the tensor product of vector spaces and $I$ is the one-dimensional space $\mathbb{R}$. In defining cups and caps, we make use of the fact that if $\{ v_i \}_i$ and $\{ u_j \}_j$ are bases for vector spaces $V$ and $U$ respectively, then $\{ v_i \otimes u_j \}_{i,j}$ is a basis for ${V\otimes U}$. Moreover, any linear map is fully determined by its action on a basis. Every finite-dimensional vector space is self-dual, and the cups and caps are given by
\begin{gather*}
\epsilon_V : V\otimes V \rightarrow \mathbb{R}::\sum_{i,j} c_{i,j} \: (v_i\otimes v_j )\mapsto \sum_{i,j} c_{i,j} \langle v_i | v_j \rangle
\\
\\
\eta_V : \mathbb{R} \rightarrow V\otimes V::1\mapsto \sum_{i} (v_i\otimes v_i )
\end{gather*}
Verifying that these maps satisfy the snake equations again follows from a straightforward calculation.
\end{example}

\begin{remark}\label{rem:collapse}\em
  The tensor in a compact closed category is \emph{not} necessarily a categorical product. Specifically, we cannot expect to describe every state of~$A \otimes B$ as a tensor
  of two states taken from~$A$ and~$B$. We can understand this as showing composite systems have interesting behaviour that cannot be explained in terms of the behaviour of their component parts.
  In fact, if the tensor of a compact closed category happens to be a categorical product, then that category must be trivial, in a precise mathematical sense~\cite{CKbook}.
\end{remark}

\subsection{Grammar categories}
Many algebraic gadgets exist to model grammar, as detailed in~\cite{Coecke2013} for example. In the present work, we use Lambek's pregroup grammars,
as many grammars have a pregroup structure~\citep{Sadrzadeh2007}. Moreover, pregroups can be viewed as compact closed categories.
It should be emphasized though that our approach does not depend on pregroups, and can be applied to other grammatical models.
\begin{dfn}
  A \define{pregroup} is a tuple $(A,\cdot , 1, \--^l,\--^r,\leq )$ where $(A,\cdot, 1, \leq)$ is a partially ordered monoid and $\--^r, \--^l$ are functions $A\rightarrow A$ such that $\forall x\in A$, 
  \begin{align}
    \label{eq:pg1}
    x\cdot x^r \leq 1\\ 
    \label{eq:pg2}
    x^l\cdot x\leq 1\\
    \label{eq:pg3}
    1 \leq x^r\cdot x\\
    \label{eq:pg4}
    1\leq x\cdot x^l
  \end{align}
\end{dfn}
The $\cdot$ will usually be omitted, writing $xy$ for $x\cdot y$. One interprets grammar by freely generating a pregroup from a set of atomic linguistic types. Words are then assigned an element of the pregroup depending on their linguistic function. A string of words is then mapped to an element of the pregroup by multiplying together the elements associated with its constituent words in their syntactic order. If $s_1\cdots s_n \leq t$, we say that the type $s_1\cdots s_n$ \define{reduces} to the type $t$. The pregroup freely generated by a set $A$ is denoted by $\freepreg A$.

\begin{example}
For simplicity, we only use the linguistic types~$n$, for noun, and~$s$, for sentence. Hence, we will work with $\freepreg{\{n,s\}}$. Consider the sentence \textit{`Chickens cross roads.'} The nouns \textit{chickens} and \textit{roads} are of type $n$, and the transitive verb \textit{cross} is assigned the type $n^r s n^l$. \textit{`Chickens cross roads'} therefore has type~$n(n^rsn^l)n$. Then we have the following type reductions:
\begin{align*}
  n(n^rsn^l)n &= (nn^r)s(n^ln)\\
  & \stackrel{(\ref{eq:pg1})}{\leq} s(n^ln) \\
  & \stackrel{(\ref{eq:pg2})}{\leq} s 
\end{align*}
Note that we could also have performed these two steps in the opposite order.
The above reduction can be given a neat graphical interpretation as follows:
\begin{center}
\input{chickens.tikz}  
\end{center}
where it is now very clear that the order of the reductions doesn't matter.
This is a feature that is typical for pregroup grammars, while other categorial grammars such as Lambek's original categorial grammar~\citep{Lambek0} have more constraints on the order of the reductions.
\end{example}

A pregroup can be considered as a compact closed category. The objects of this category are the elements of the pregroup. The morphisms are given by the order structure of the pregroup. That is, there is a unique morphism~$p\rightarrow q$ if and only if~$p\leq q$. The  tensor~$\otimes$ is the monoid multiplication and the monoidal unit is the element $1$.
Unsurprisingly, the left and right duals of~$p$ are~$p^l$ and~$p^r$ respectively.
The cups and caps are the unique morphisms given by the pregroup axioms~(\ref{eq:pg1})--(\ref{eq:pg4}).
In the reduction diagram, note how the cups correspond to the cups of the compact closed structure.
\subsection{Meaning categories}
Distributional models of meaning use vector spaces to represent the meaning of words. In this paper we move away from this approach, choosing instead to work in \convexrel , the category of convex sets and convexity-respecting relations. Before describing this new setting, we first present the conventional vector space approach as a demonstration of the categorical compositional method. This will prepare the ground for detailed discussion of~\convexrel to later sections.

One way to model meanings in a vector space is to use co-occurrence statistics~\citep{Bullinaria2007}. The meaning of a word is identified with the frequency with which it appears near other words. One first chooses a collection of \textit{context words}. These will be the basis vectors. One then analyses a large corpus of writing to determine how often words co-occur with the context words. For example, suppose the word \textit{dog} appears in the same context as \textit{cat}, \textit{companion}, and \textit{cuisine} respectively 19, 25, and 2 times. If our collection of context words is $\{ \mbox{\it cat}, \mbox{\it companion}, \mbox{\it cuisine} \}$, then \textit{dog} would be assigned the vector $(19,25,2)$. A drawback of the co-occurrence approach is that antonyms appear in similar contexts and, hence, words such as `win' and `lose' are indistinguishable (despite evidently being different in meaning). Another related difficulty is that vector spaces are notoriously bad for representing basic propositional logic.  Nonetheless, the vector space model is highly successful in NLP.

One can also use basis vectors to represent \textit{quality dimensions}. For example, \cite{rosch1975, tversky1977, hampton1987} all represent concepts as feature vectors, with basis dimensions representing attributes of the concept. So, for instance, one might represent the word \textit{dog} with certain values on the \textit{fluffy}, \textit{loyal}, and \textit{wolf-like}. This approach closely resembles ours in this paper, though we move away from the vector space setting.

\subsection{Putting it all together}
We now have all the necessary components: a grammar category and a meaning category. We used the examples of pregroup grammars and vector spaces, but we stress yet again
that the abstract method applies equally well to any two compact closed categories.

Suppose we have a string of words $w_1\cdots w_n$ and a pregroup $P$. Suppose further that $w_i\in W_i$, where $W_i$ is the vector space associated with $w_i$'s linguistic type.  The meaning of $w_1\cdots w_n$ is computed as follows:
\begin{enumerate}
\item Assign a pregroup element $p_i$ to each word $w_i$ based on its linguistic type.
\item Apply pregroup reduction rules (cups and caps) to the element $p_1p_2\cdots p_n$ to obtain a simpler type $x$ such that: \[
p_1p_2\cdots p_n\leq x.
\]
\item Thinking of the above reduction as a morphism in the pregroup built up from $\epsilon, \eta$ and identities (as given by its reduction diagram), apply the corresponding vector space morphism of type: 
\[
W_1\otimes \cdots \otimes W_n \rightarrow W_X 
\]
to the string of word meanings represented as~$w_1\otimes \cdots \otimes w_n$.
\end{enumerate}
\begin{example}
Consider again the sentence \textit{`chickens cross roads'}. The nouns \textit{chickens} and \textit{roads} have type $n$ and so are represented in some vector space $N$ of nouns. The transitive verb \textit{cross} has type $n^r s n^l$ and, hence, is represented by a vector in the vector space $N\otimes S \otimes N$ where $S$ is a vector space modelling sentence meaning. The meaning of \textit{`chickens cross roads'} is the image of
\begin{equation}\label{eq:pgex1}
\vect{\mathit{chickens}}\otimes\vect{\mathit{cross}}\otimes\vect{\mathit{roads}}
\end{equation}
under the map
\begin{equation}\label{eq:pgex2}
\epsilon_N \otimes 1_S \otimes \epsilon_N : N\otimes ( N\otimes S\otimes N ) \otimes N \rightarrow S
\end{equation}
\end{example}
This nicely illustrates the general method. Our meaning category supplies the qualitative meanings of \textit{chickens}, \textit{cross}, and \textit{roads}. Our grammar category then tells us how to stitch these together. This corresponds to `telling us where to put cups and caps.' The essence of the method should be thought of as the diagram
\begin{center}
\input{chickens2.tikz}
\end{center}
where we think of the words as meaning vectors (\ref{eq:pgex1}) and the wires as the map (\ref{eq:pgex2}).  In fact, rather than just using cups and identity wires, which are enough to account for the grammatical structure captured by pregroups,  we can enrich the graphical language to directly account for meanings of functional words, such as relative pronouns.   

\subsection{Beyond standard categorial grammar}
A first rather trivial example of a functional word is ``does", which can be accounted for by means of caps \cite{CoeckeSadrzadehClark2010}:
\begin{center}
\input{chickens3.tikz}
\end{center}
Things become more interesting when we introduce, rather than just wires, the idea of a \em multi-wire \em (also called \em spider \em \citep{CKbook}) which can have more than two ends, or fewer:
\[
\mu_N := \begin{gathered}\input{spider2a.tikz}\end{gathered} \qquad\qquad\qquad\qquad\qquad\qquad \iota_S := \begin{gathered}\input{spider2b.tikz}\end{gathered}
\]
The way these behave is just like wires.  The only thing that matters is: `either being connected by a multi-wire, or not'.  As a consequence, multi-wires `fuse' together: 
\begin{center}
\input{spider1.tikz}
\end{center}
These multi-wires can be defined in category-theoretic terms for any symmetric monoidal category, where they are called \em commutative special dagger Frobenius structures \em \citep{CPV, CKbook}.

\begin{example}
On each set $X$ in the category $\mathbf{Rel}$ one can take the relations
\[
 X\otimes \ldots \otimes  X\rightarrow X\otimes \ldots \otimes  X :: \{ \left(( x, \ldots ,x), ( x, \ldots ,x)\right) \mid x \in X \}
\]
to be the multi-wires, and note in particular that these include identities, cups and caps.
\end{example}

\begin{example}
On each inner product space $V$ in the category $\mathbf{FdVect}_\mathbb{R}$, given any orthonormal basis $\{ v_i \}_i$ for $V$, one can take the linear maps
\[
V\otimes \ldots \otimes  V\rightarrow V\otimes \ldots \otimes  V :: v_{i_1}\otimes \ldots \otimes v_{i_n}\mapsto \delta_{i_1\ldots i_n} v_{i_1}\otimes \ldots \otimes v_{i_1}
\]
to be the multi-wires, and again these include identities, cups and caps.
\end{example}

Using these multi-wires we can now express the meaning of relative pronouns \citep{FrobMeanI, FrobMeanII}:
\begin{center}
\input{chickens4.tikz}
\end{center}
Firstly, note that what we obtain is a noun rather than a sentence, and one that is closely related to `dead chicken'.  Secondly, note that the use of the three-wire is mainly conjunction, conjoining `[is] chicken' and `crosses road'.  In this paper we will use them to directly express conjunctions.   The one-wire gets rid of the sentence type.

\section{Conceptual spaces}
\emph{Conceptual spaces} are proposed in \cite{Gardenfors2004} as a framework for representing information at the conceptual level. G\"{a}rdenfors contrasts his theory with both a symbolic approach to concepts, and an associationist approach where concepts are represented as associations between different kinds of information elements. Instead, conceptual spaces are structures based on quality dimensions such as weight, height, hue and brightness. Conceptual spaces have an internal structure based on how quality dimensions interact with each other. A pair (or set) of dimensions is called \emph{integral} if assignment of a value on one dimension requires assignment of a value on another dimension. For example, the dimensions of hue, saturation, and value in the HSV colour space are integral. Dimensions are called \emph{separable} if values on one dimension can be assigned independently from the others. For example, hue and height are separable. A set of integral dimensions is called a domain, and a conceptual space is composed of a number of domains linked in some way. A concept then corresponds to a convex region in a conceptual space. The way in which the interaction between quality dimensions is specified in G\"ardenfors's model is by using different distance metrics. Within a set of integral dimensions, distance is Euclidean, and between sets of integral dimensions, the city block metric is used.

Concept composition within conceptual spaces has been formalized in~\cite{RickardAisbettGibbon2007, AdamsRaubal2009,  LewisLawry2016} for example. All these approaches focus on noun-noun composition, rather than utilising any more complex structure, and the way in which nouns compose often focuses on correlations between attributes in concepts. Since then, G\"ardenfors has started to formalise verb spaces, adjectives, and other linguistic structures \citep{Gardenfors2014}. However, a systematic method for how to utilise grammatical structures within conceptual spaces has not yet been provided. In this sense, the category-theoretic approach to concept composition we describe below will introduce a more general approach to concept composition that can apply to varying grammatical types.

\section{The category of convex relations}
\label{sec:convexrel}
In NLP applications, meanings are typically interpreted in categories of real vector spaces.
For our intended cognitive application, we now introduce a category
that emphasizes convex structure. The familiar definition of convex set is a subset of a vector space which is closed under
forming convex combinations. In this paper we consider a more general setting that includes
convex subsets of vector spaces, but also allows us to consider some further, more discrete, examples.

We begin with some convenient notation. For a set~$X$ we write~$\sum_i p_i \ket{ x_i }$
for a finite formal convex sum of elements of~$X$, where $p_i \in \mathbb{R}^{\geq 0}$ and~$\sum_i p_i = 1$. We then write~$D(X)$ for the set of all such sums.
Here we abuse the physicists ket notation
to highlight that our sums are formal, following
a convention introduced in~\cite{Jacobs2011b}.
Equivalently, these sums can be thought of as finite probability distributions on the elements of~$X$.

A \define{convex algebra} is a set~$A$ with a function~$\alpha : D(A) \rightarrow A$ satisfying the following conditions:
\begin{equation}
  \label{eq:convexalg}
  \alpha(\ket{ a }) = a\qquad\text{ and }\qquad \alpha\left(\sum_{i,j} p_i q_{i,j} \ket{ a_{i,j} }\right) = \alpha\left(\sum_i p_i \ket{ \alpha(\sum_j q_{i,j} \ket{ a_{i,j} }) }\right)
\end{equation}
Informally, $\alpha$ is a \define{mixing operation} that allows us to form convex combinations of elements,
and the equations in~\eqref{eq:convexalg} require the following good behaviour:
\begin{itemize}
\item Forming a convex combination of a single element~$a$ returns~$a$ as we would expect
\item Iterating forming convex combinations interacts with flattening sums of sums in the way we would expect
\end{itemize}
We consider some examples of convex algebras.
\begin{example}
  The closed real interval~$[0,1]$ has an obvious convex algebra structure.
  Similarly, every real or complex vector space has a natural convex algebra structure using the underlying linear structure.
\end{example}
\begin{example}[Simplices]
  For any set~$X$, the formal convex sums of elements of~$X$ themselves form the \define{free convex algebra} on~$X$, which
  can also be seen as a simplex with vertices the elements of~$X$. Mixtures are formed as follows:
  \begin{equation*}
    \sum_i p_i \ket{ \sum_j q_{i,j} \ket{ x_{i,j} } } \mapsto \sum_{i,j} p_i q_{i,j} \ket{ x_{i,j} }
  \end{equation*}
\end{example}
\begin{example}
  The convex space of density matrices provides another example, with the convex structure given by
  the usual vector space structure on linear operators.
\end{example}
\begin{example}
  \label{ex:fuzzy}
  For a set $X$, the functions of type $X \rightarrow [0,1]$ form a convex algebra pointwise,
  with mixing operation:
  \begin{equation*}
    \sum_i p_i \ket{ f_i } \mapsto (\lambda x. \sum_i p_i f_i(x))
  \end{equation*}
  We can see this as a convex algebra of fuzzy sets.
\end{example}
\begin{example}[Semilattices]
  \label{ex:semilattices}
  As a slightly less straightforward example, every affine
  join semilattice (that is, one that has all finite \emph{non-empty} joins) has a convex algebra structure given by:
  \begin{equation*}
    \sum_i p_i \ket{ a_i } = \bigvee_i \{ a_i \mid p_i > 0 \}
  \end{equation*}
  Notice that here the scalars~$p_i$ are discarded and play no active role. These ``discrete'' types
  of convex algebras allow us to consider objects such as the Boolean truth values.
\end{example}
\begin{example}[Trees]
  \label{ex:concretesl}
  Given a finite tree, perhaps describing some hierarchical structure, we can construct an affine semilattice in a natural way.
For example, consider a limited universe of foods, consisting of bananas, apples, and beer. Given two members of the hierarchy, their join will be the lowest level of the hierarchy which is above them both. For instance, the join of $\lang{bananas}$ and $\lang{apples}$ would be $\lang{fruit}$.
\begin{equation*}
\Tree
    [.\lang{food} [.\lang{fruit} \lang{apples} \lang{bananas} ] [.\lang{beer} ] ]
\end{equation*}
\end{example}

When~$\alpha$ can be understood from the context, we abbreviate our notation for convex combinations by writing:
\begin{equation*}
  \sum_i p_i a_i := \alpha(\sum_i p_i \ket{ a_i })
\end{equation*}
Using this convention, we define a \define{convex relation} of type~$(A,\alpha) \rightarrow (B,\beta)$ as
a binary relation $R : A \rightarrow B$ between the underlying sets that commutes with forming convex mixtures as follows:
\begin{equation*}
  \left(\forall i. R(a_i, b_i)\right) \Rightarrow R\left(\sum_i p_i a_i, \sum_i p_i b_i\right)
\end{equation*}
We note that identity relations are convex, and convex relations are closed under relational composition and converse.
\begin{example}[Homomorphisms]
  If~$(A,\alpha)$ and~$(B, \beta)$ are convex algebras, functions $f:A \rightarrow B$ satisfying:
  \begin{equation*}
    f(\sum_i p_i x_i) = \sum_i p_i f(x_i)
  \end{equation*}
  are convex relations. These functions are the \define{homomorphisms of convex algebras}.
  The identity function and constant functions are examples of homomorphisms of convex algebras.
\end{example}
The singleton set $\{ * \}$ has a unique convex algebra structure, denoted~$I$.
Convex relations of the form~$I \rightarrow (A,\alpha)$ correspond to~\define{convex subsets},
that is, subsets of~$A$ closed under forming convex combinations.
\begin{dfn}
  We define the category~\convexrel as having convex algebras as objects and convex relations
  as morphisms, with composition and identities as for ordinary binary relations.
\end{dfn}

Given a pair of convex algebras $(A,\alpha)$ and~$(B,\beta)$ we can form a new convex algebra
on the cartesian product $A \times B$, denoted~$(A,\alpha) \otimes (B,\beta)$,
with mixing operation:
\begin{equation*}
  \sum_i p_i \ket{ (a_i, b_i) } \mapsto \left(\sum_i p_i a_i, \sum_i p_i b_i\right)
\end{equation*}
This induces a symmetric monoidal structure on~\convexrel.
In fact, the category~\convexrel has the necessary categorical structure for categorical~compositional~semantics:
\begin{thm}
  \label{thm:convexrel}
  The category~\convexrel is a
  compact closed category.
  The symmetric monoidal structure is given by the unit and monoidal product outlined above.
  The caps for an object $(A,\alpha)$ are given by:
  \[
  \raisebox{-0.2cm}{\input{cap.tikz}}
  : I \rightarrow (A,\alpha) \otimes (A,\alpha)::
  \{ (*,(a,a)) \mid a \in A \}
  \]
  the cups by: 
  \[
  \raisebox{-0.2cm}{\input{cup2.tikz}}
  :(A,\alpha) \otimes (A,\alpha) \rightarrow I :: \{ ((a,a),*) \mid a \in A \}
  \]
  and more generally, the multi-wires by:
  \[
  \raisebox{-0.7cm}{\input{spider.tikz}}: A\otimes \ldots \otimes  A\rightarrow A\otimes \ldots \otimes  A :: \{ \left(( a, \ldots ,a), ( a, \ldots ,a)\right) \mid a \in A \}
  \]
\end{thm}

Note that in the definition of the multi-wires and for the remainder of the paper, we abuse notation and leave the algebra $\alpha$ on $A$ implicit.

\begin{remark}\em
\label{rem:mergedel}
In particular, the multi-wires $\mu_A$ and $\iota_A$ are defined as follows:
\[
\mu_A:A \otimes A \rightarrow A ::\{((a, a), a)|a \in A\}
\qquad
\iota_A: A \rightarrow I :: \{(a, *)| a \in A\}
\]
\end{remark}

\begin{remark}\em
  As observed in remark~\ref{rem:collapse}, as~\convexrel is compact closed, its tensor cannot be a categorical product.
  For example, there are convex subsets of~$[0,1]\otimes[0,1]$ such as the diagonal:
  \begin{equation*}
    \{ (x,x) \mid x \in [0,1] \}
  \end{equation*}
  that cannot be written as the cartesian product of two convex subsets of~$[0,1$].
  This behaviour exhibits non-trivial \emph{correlations} between the different components of the composite convex algebra. 
\end{remark}

\begin{remark}\em
We have given an elementary description of~\convexrel. More abstractly, it can be seen as the category
of relations in the Eilenberg-Moore category of the finite distribution monad. Its compact closed
structure then follows from general principles~\citep{CarboniWalters1987}.
\end{remark}

\section{Noun, adjective, and verb concepts}
We define a~\define{conceptual space} to be an object of~\convexrel.
In order to match the structure of the pregroup grammar, we require two distinct objects: a noun space~$N$ and a sentence space~$S$.

The \define{noun~space}~$N$ is given by a composite
\[
N_\lang{colour} \otimes N_\lang{taste} \otimes ...
\]
describing different attributes such as colour and taste.
A~\define{noun} is then a convex subset of such a space. 
In our examples, we take our sentence space to be a convex algebra in which the individual points are events. Our general scheme can
incorporate other sentence space structures, such choices are generally specific to the application under consideration.
A~\define{sentence} is then a convex subset of~$S$.

We now describe some example noun and sentence spaces. We then show how these can be combined to form spaces describing adjectives and verbs.
Once we have these types available, we show in section~\ref{sec:composing} how concepts interact within sentences.

\subsection{Example: Food and drink}
\begin{figure}
\begin{subfigure}[t]{0.3\textwidth}
\includegraphics[width =\textwidth]{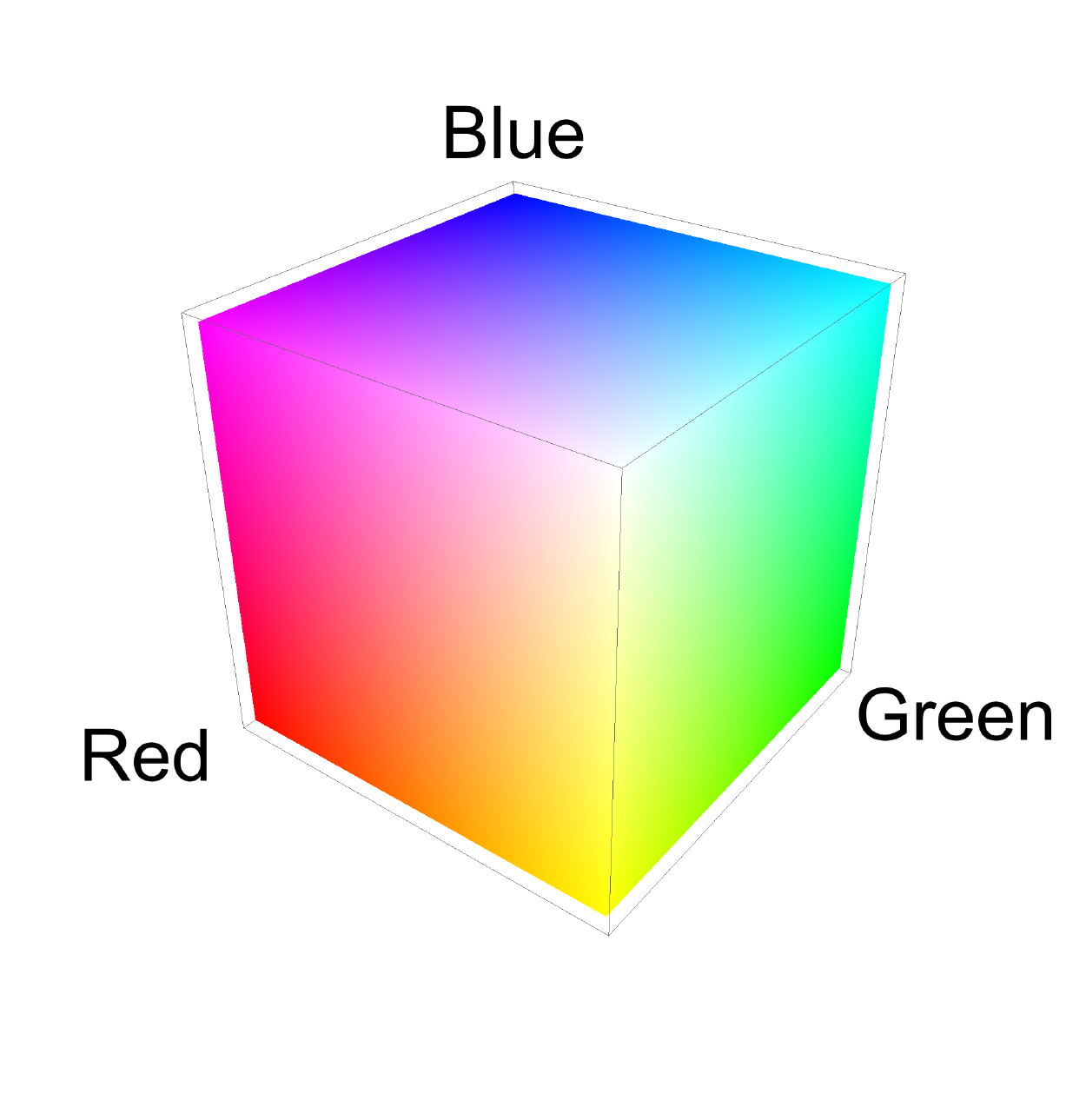}
\caption{The RGB colour cube}
\end{subfigure}
\hfill
\begin{subfigure}[t]{0.3\textwidth}
\includegraphics[width = \textwidth]{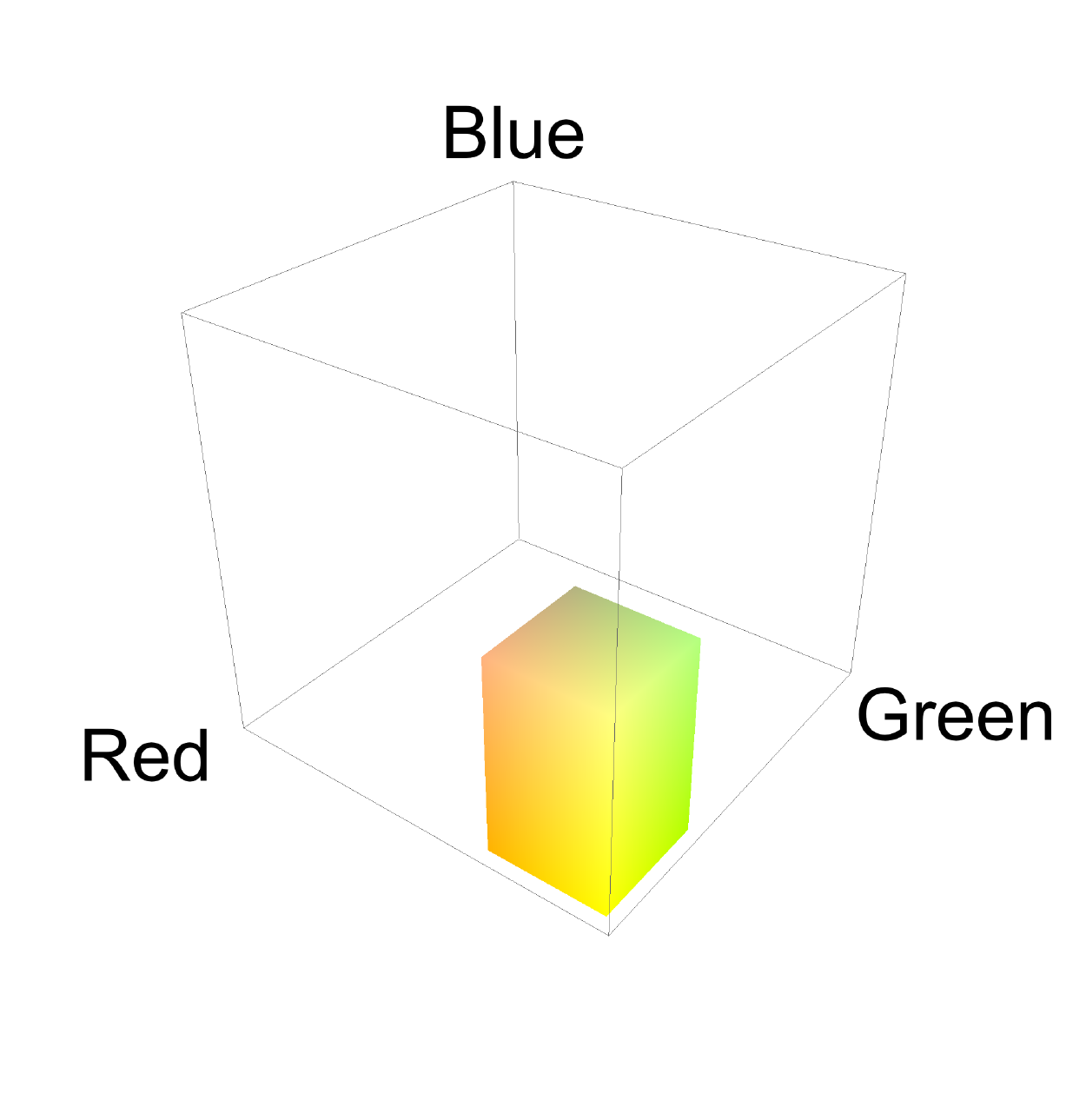}
\caption{Property $\property{yellow}$}
\end{subfigure}
\hfill
\begin{subfigure}[t]{0.3\textwidth}
\includegraphics[width = \textwidth]{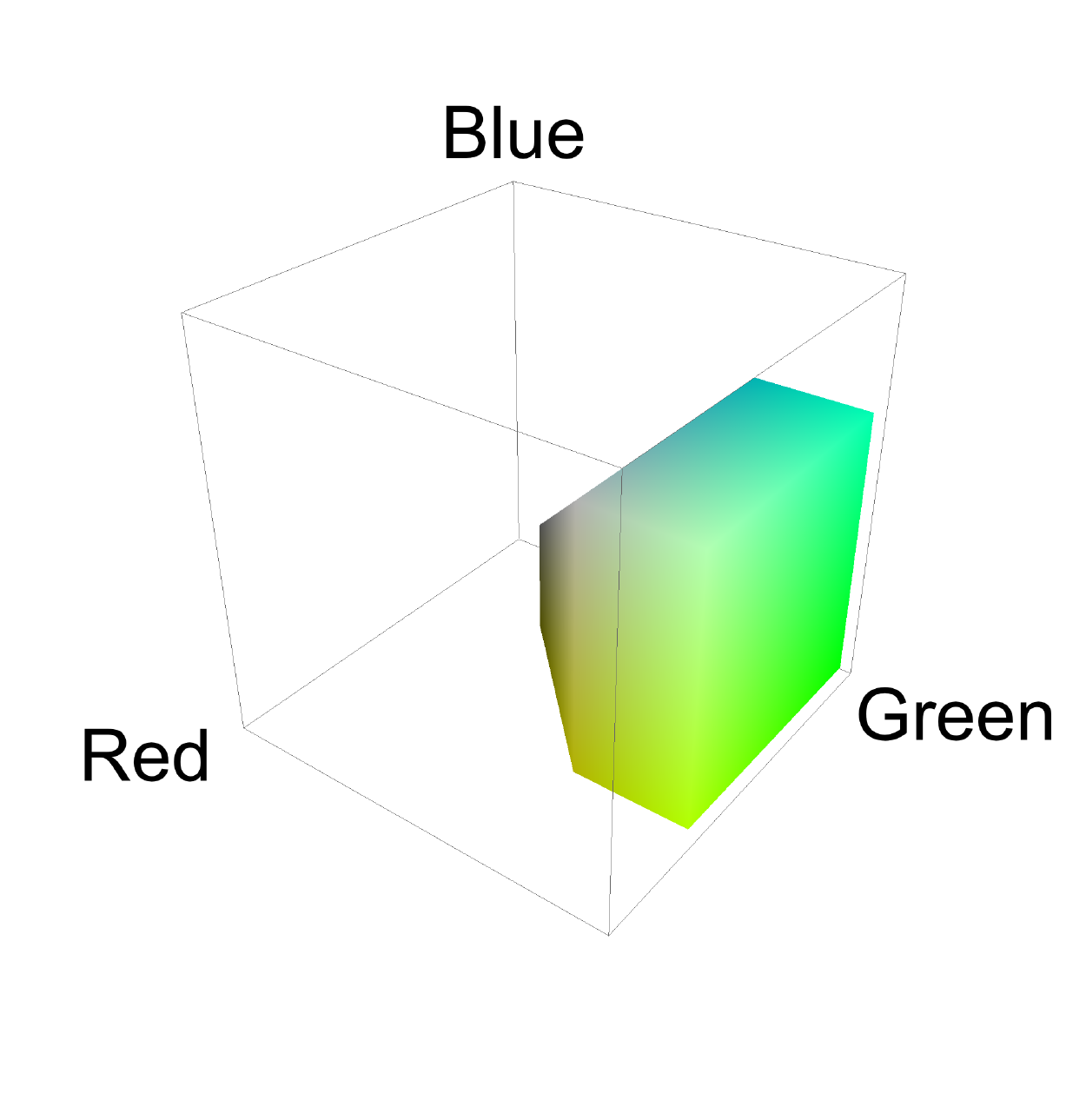}
\caption{Property $\property{green}$}
\end{subfigure}
\caption{The RGB colour cube and properties $\property{colour}$}
\label{fig:colour_space}
\end{figure}

\begin{figure}
\centering
\begin{subfigure}[t]{0.35\textwidth}
\includegraphics[width = \textwidth]{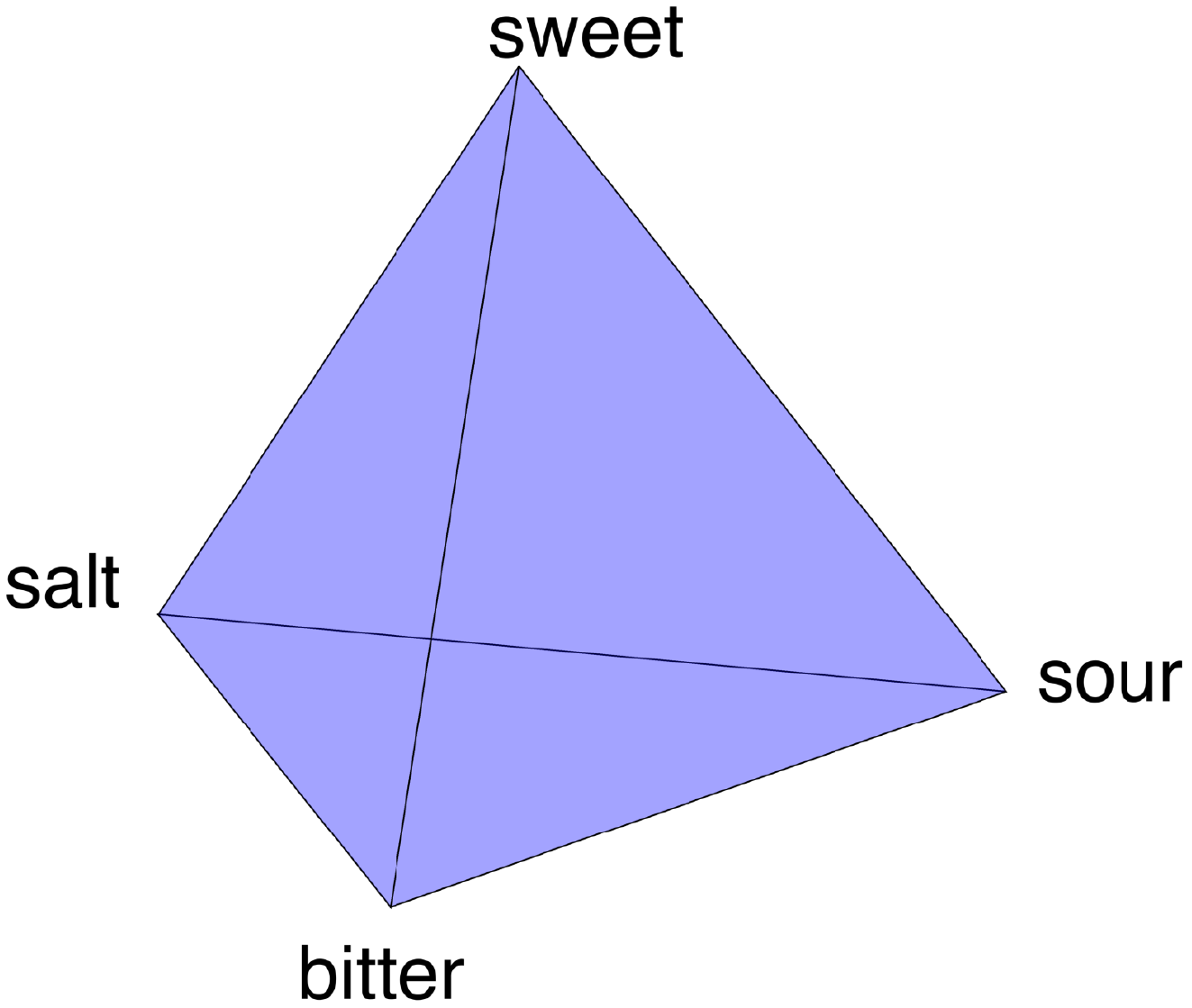}
\caption{The taste tetrahedron}
\end{subfigure}
\hfill
\begin{subfigure}[t]{0.35\textwidth}
\includegraphics[width = \textwidth]{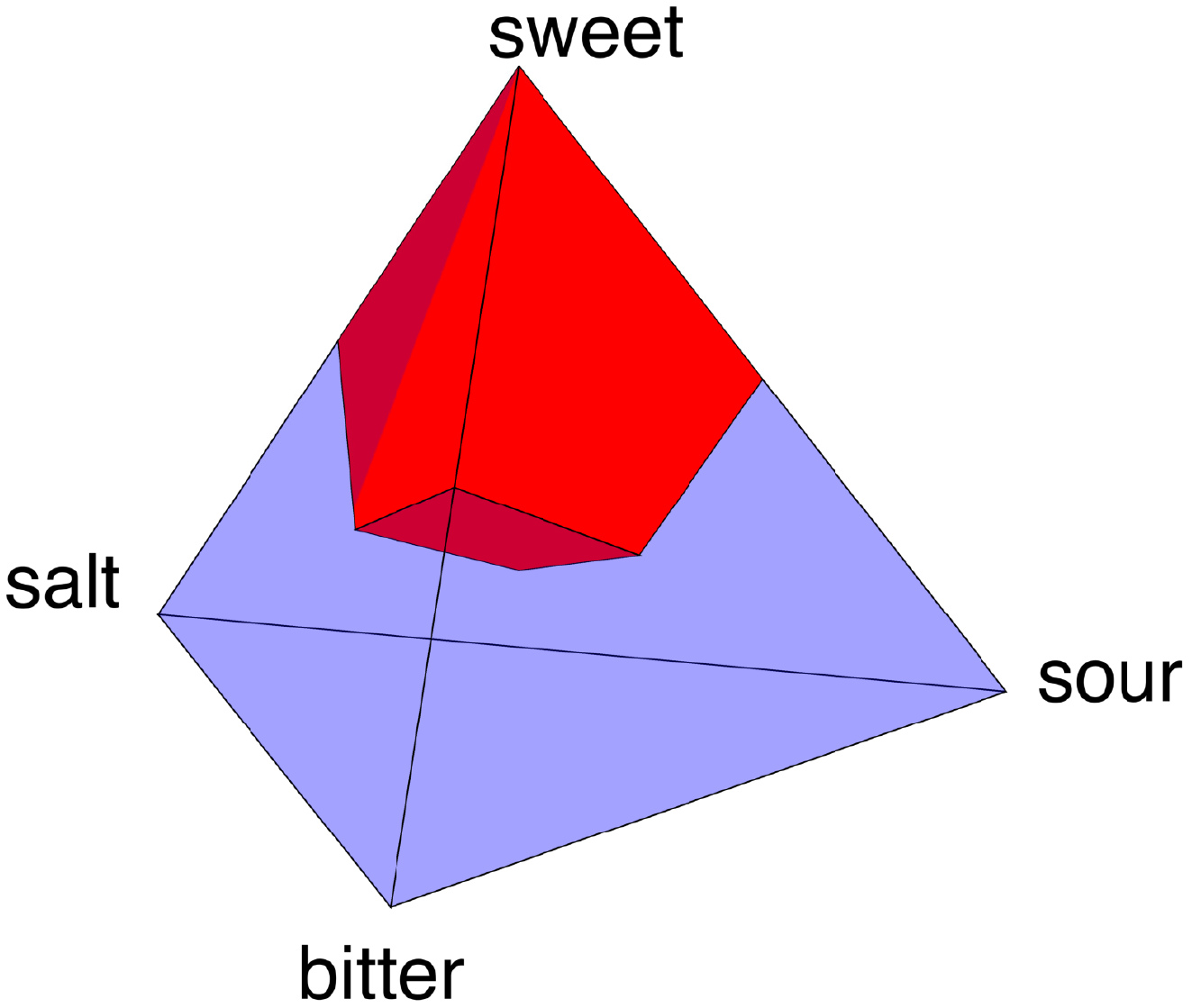}
\caption{The region corresponding to $\property{sweet} = \{\vec{t}| t_\lang{sweet} \geq t_l \text{ for } l \neq \lang{sweet}\}$}
\end{subfigure}
\caption{The taste space and the property \property{sweet}}
\label{fig:taste_space}
\end{figure}

We consider a conceptual space for food and drink as our running example. The space~$N$ is composed of the domains $N_\lang{colour}$, $N_\lang{taste}$, $N_\lang{texture}$, so that 
\[
N = N_\lang{colour} \otimes N_\lang{taste} \otimes N_\lang{texture}
\]
The domain~$N_\lang{colour}$ is the RGB colour domain, i.e. triples $(R, G, B) \in [0,1]^3$ with $R$, $G$, $B$ standing for intensity of red, green, and blue light respectively. $N_\lang{taste}$ is defined as the simplex of combinations of four tastes: sweet, sour, bitter, and salt. We therefore have
\begin{equation}
\label{eq:taste_space}
N_\lang{taste} = \{\vec{t} | \vec{t} = \sum_{i \in I} w_i \vec{t}_i\}
\end{equation}
where  $I = \{\lang{sweet}, \lang{sour}, \lang{bitter}, \lang{salt}\}$, $\vec{t}_i$ is the vector in some chosen basis of $\mathbb{R}^4$ whose elements are all zero except for the  $i$th element whose value is one, and $\sum_i w_i = 1$. $N_\lang{texture}$ is just the set $[0, 1]$ ranging from completely liquid (0) to completely solid (1). 
We define a property \property{property} to be a convex subset of a domain, and specify the following examples (see figures \ref{fig:colour_space}  and \ref{fig:taste_space}):
\begin{align*}
\property{yellow} &= \{(R, G, B)|(R \geq 0.7), (G \geq 0.7), (B \leq 0.5)\}\\
\property{green} &= \{(R, G,B)|(R \leq G), (B \leq G), (R \leq 0.7), (B \leq 0.7), (G \geq 0.3)\}\\
\property{sweet} &= \{\vec{t}| t_\lang{sweet} \geq t_l \text{ for } l \neq \lang{sweet}\}
\end{align*}
The properties~$\property{sour}$ and~$\property{bitter}$ are defined analogously. 

\subsubsection{Nouns}
We define some nouns below. Properties in the colour domain are specified using sets of linear inequalities, and colours in the taste domain are specified using the convex hull of sets of points. We use $\ch{A}$ to refer to the convex hull of a set $A$.
\begin{align*}
\lang{banana} &= \{(R, G, B)|(0.9 R \leq G \leq 1.5 R), (R \geq 0.3), (B\leq 0.1)\} \\&\qquad \otimes \ch{\{t_\lang{sweet}, 0.25t_\lang{sweet}+0.75t_\lang{bitter}, 0.7t_\lang{sweet}+0.3t_\lang{sour}\}} \otimes [0.2, 0.5]\\
\lang{apple} &= \{(R, G, B)|R - 0.7 \leq G \leq R + 0.7), (G \geq 1-R), (B\leq 0.1)\} \\&\qquad \otimes [0.5, 1] \otimes \ch{\{t_\lang{sweet}, 0.75t_\lang{sweet}+0.25t_\lang{bitter}, 0.3t_\lang{sweet}+0.7t_\lang{sour}\}} \otimes [0.5, 0.8]\\
\lang{beer} &= \{(R, G, B)|(0.5 R \leq G \leq R), (G \leq 1.5 - 0.8R), (B\leq 0.1)\} \\&\qquad \otimes \ch{\{t_\lang{bitter}, 0.7t_\lang{sweet}+0.3t_\lang{bitter}, 0.6t_\lang{sour}+0.4t_\lang{bitter}\}} \otimes [0, 0.01]
\end{align*}
where $t_i$ are as given in \eqref{eq:taste_space}. The tensor product $\otimes$ used in these equations is the tensor product in \convexrel, and is therefore the Cartesian product of sets.

The subsets of points representing tastes are explained as follows using the case of banana as an example. Bananas are not at all salty, and therefore $w_\lang{salt}$ is set to $0$. Bananas are sweet, and therefore the point $t_\lang{sweet}$ is chosen as an extremal point in the set of banana tastes. Bananas can also be somewhat but not totally bitter, and therefore the point $0.25t_\lang{sweet}+0.75t_\lang{bitter}$ is chosen as an extremal point. Similarly bananas can be a little sour, and therefore $0.7t_\lang{sweet}+0.3t_\lang{sour}$ is also chosen as an extremal point. Finally the convex hull of these points is formed giving a set of points corresponding to banana taste.

Pictorially, we have:
\begin{align*}
\lang{banana} \ \ &=\ \  \begin{gathered} \includegraphics[width=0.2\textwidth]{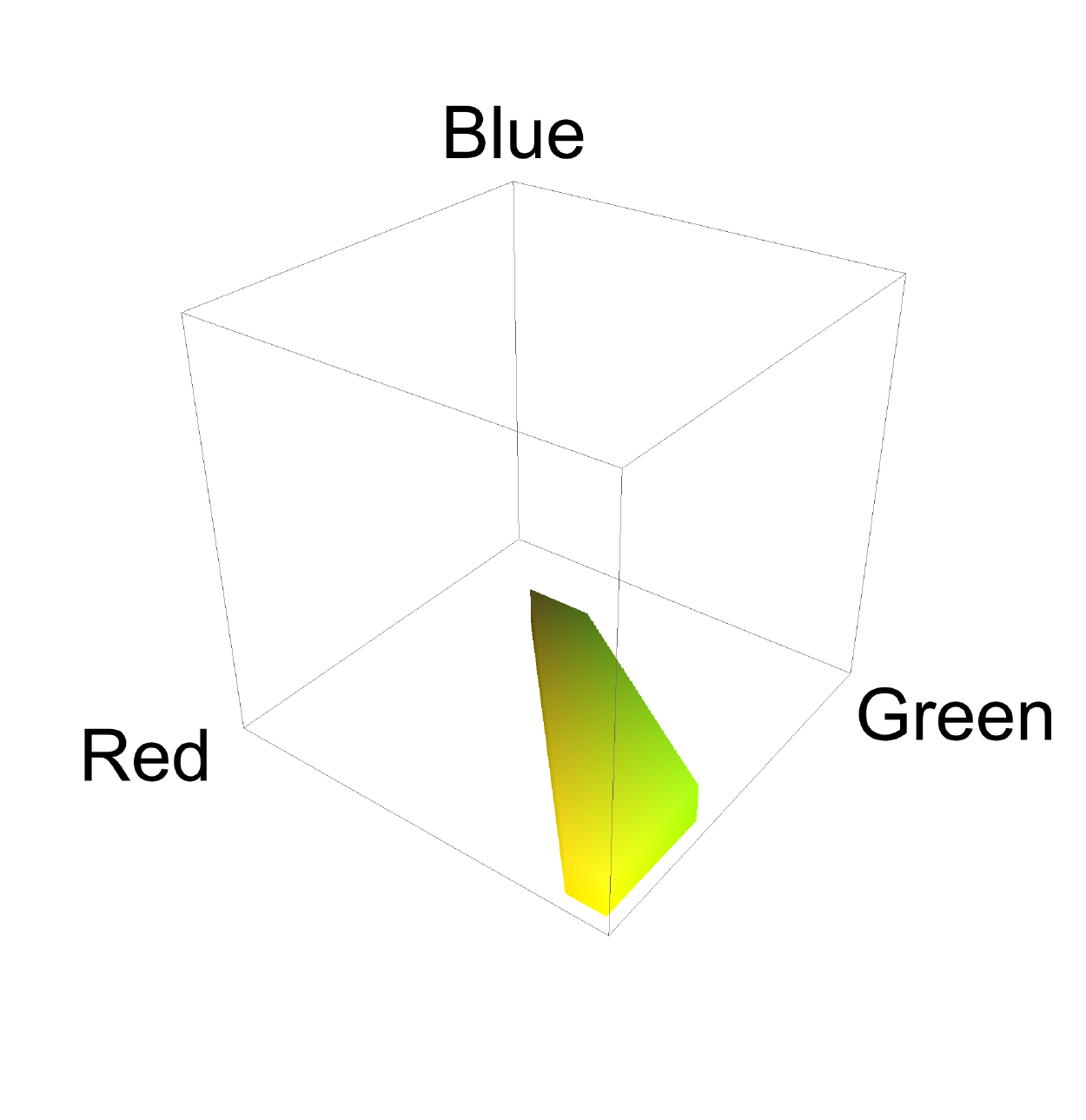}\end{gathered} \ \ \otimes\ \  \begin{gathered} \includegraphics[width=0.2\textwidth]{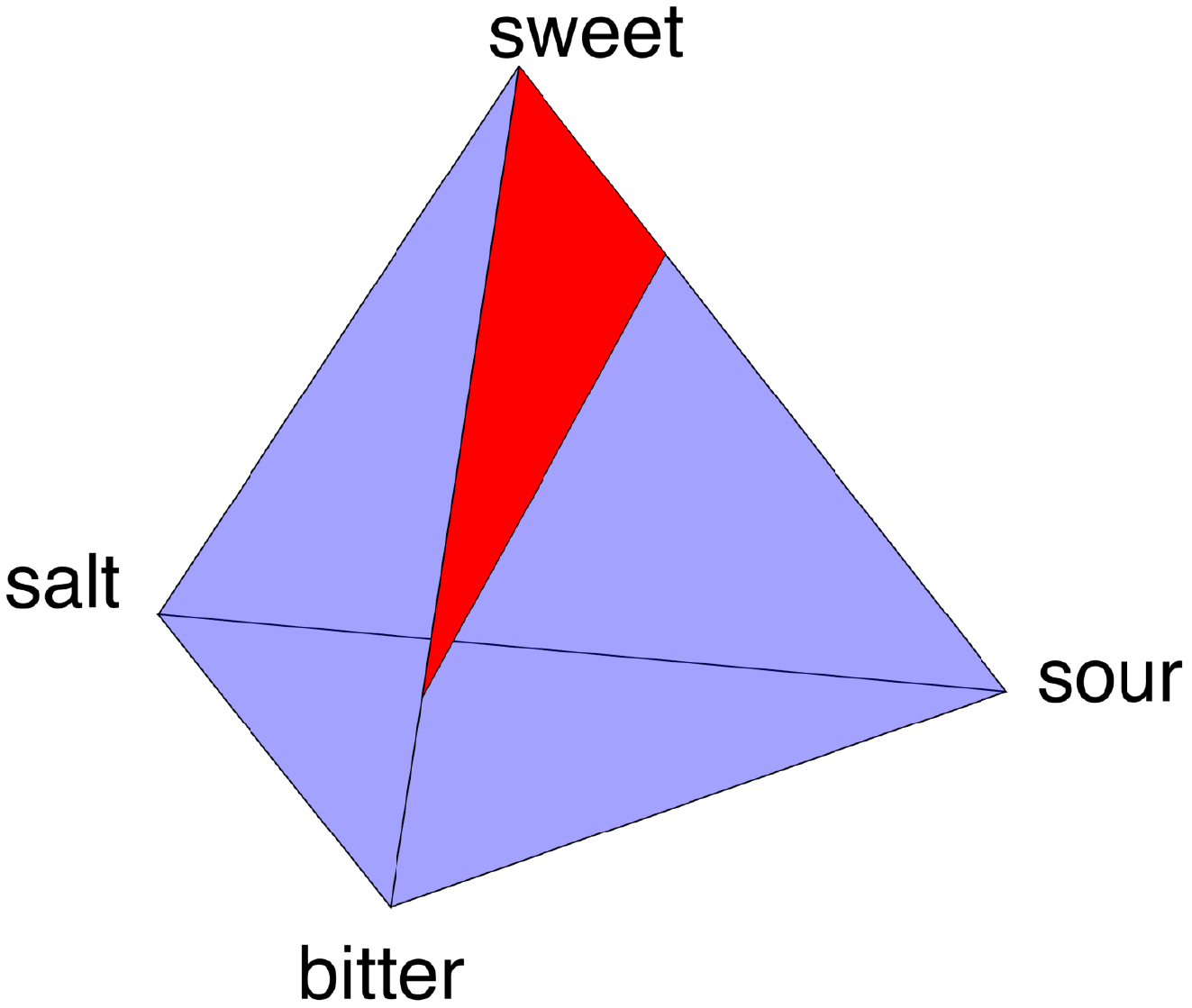}\end{gathered}\ \ \otimes\ \ \begin{gathered} \includegraphics[width=0.2\textwidth]{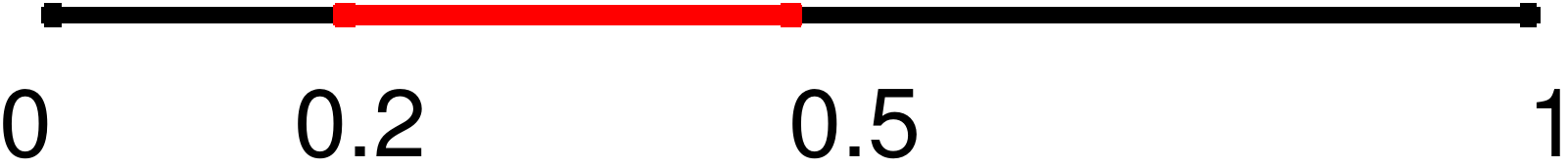}\end{gathered}\\
\lang{apple} \ \ &=\ \  \begin{gathered} \includegraphics[width=0.2\textwidth]{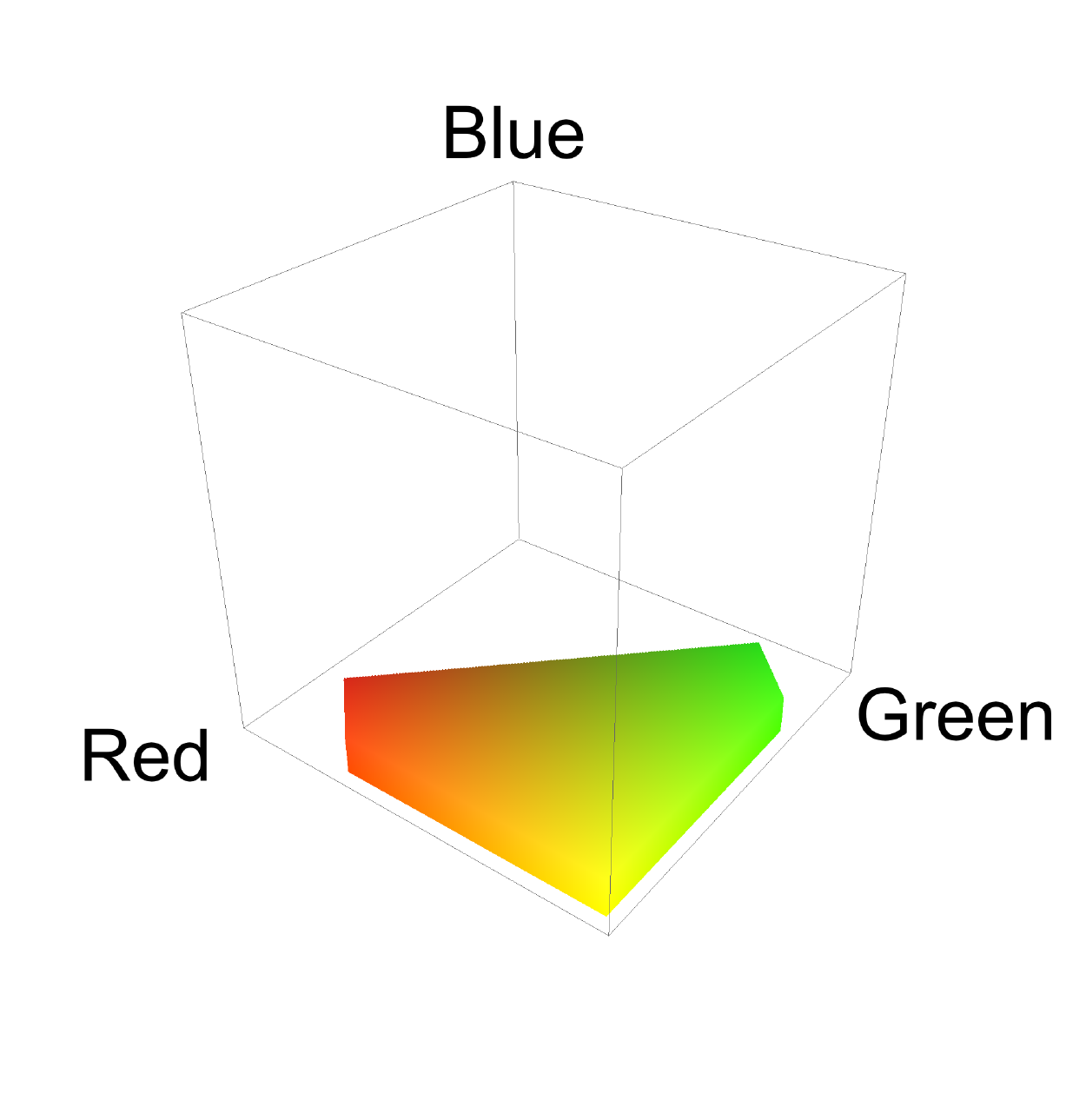}\end{gathered} \ \ \otimes\ \  \begin{gathered} \includegraphics[width=0.2\textwidth]{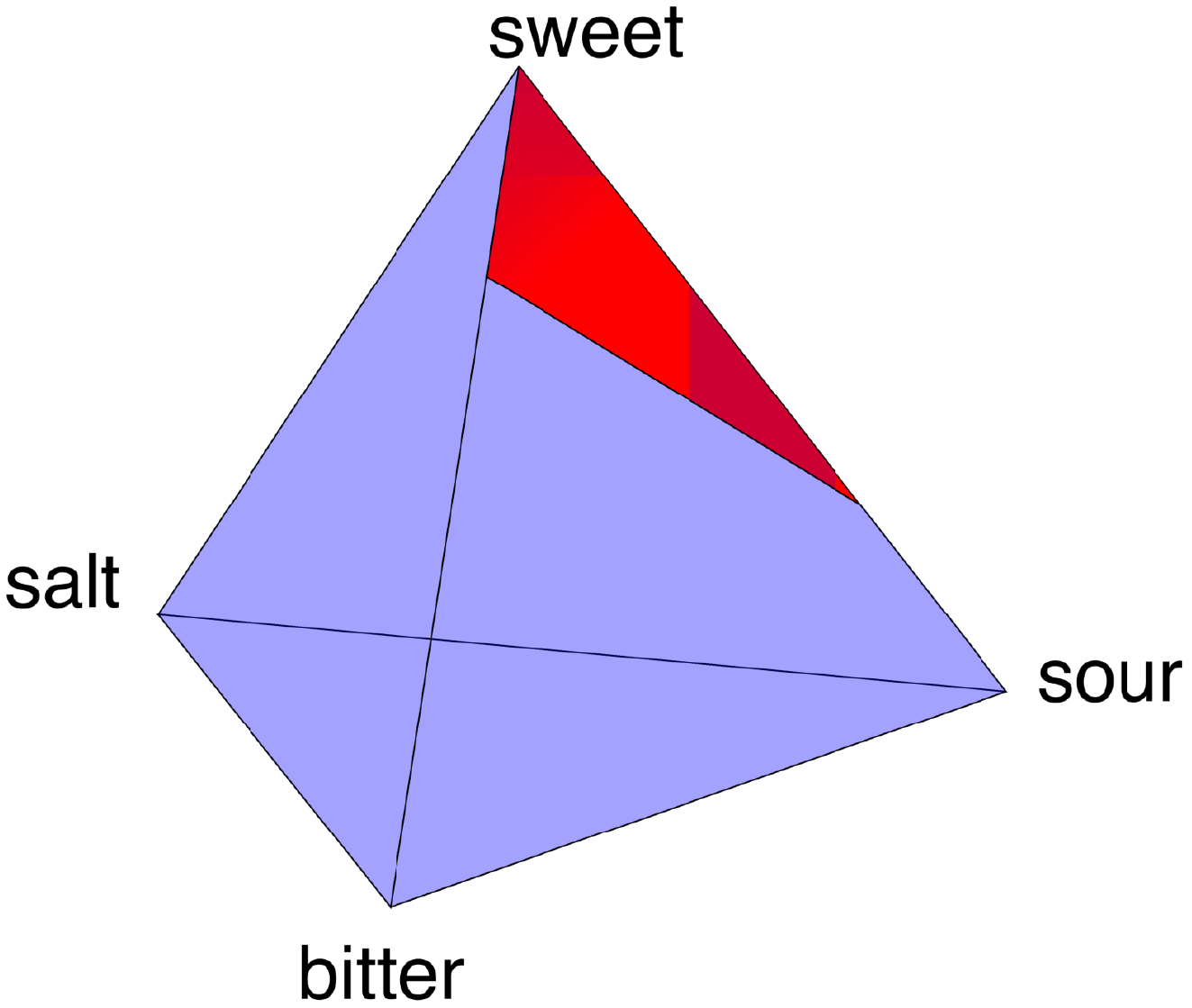}\end{gathered}\ \ \otimes\ \ \begin{gathered} \includegraphics[width=0.2\textwidth]{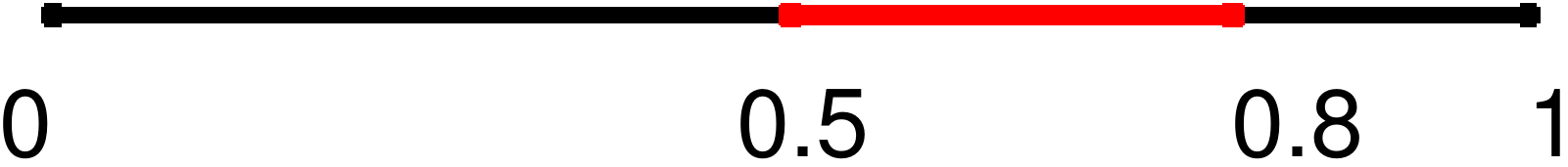}\end{gathered}\\
\lang{beer} \ \ &=\ \  \begin{gathered} \includegraphics[width=0.2\textwidth]{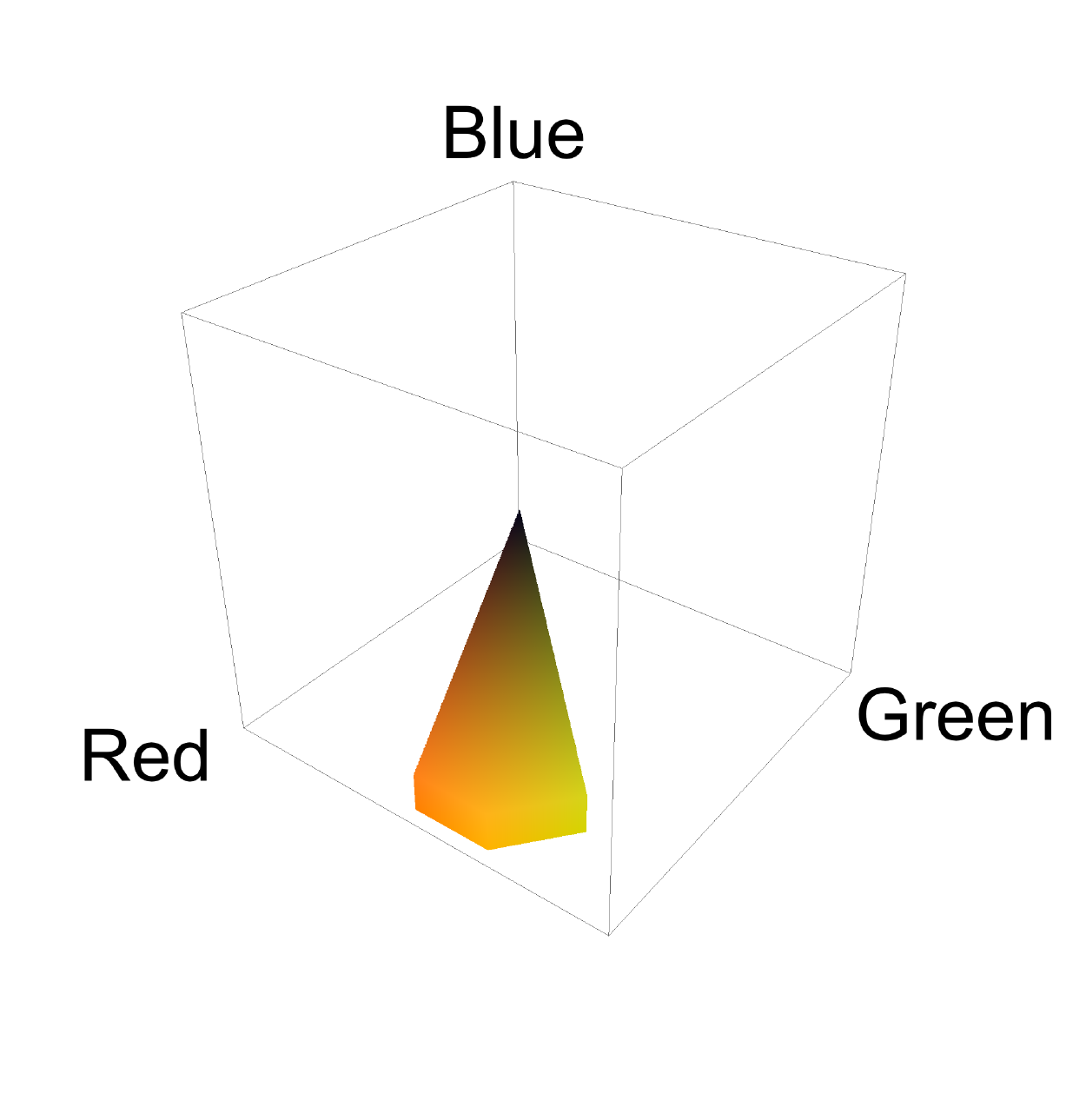}\end{gathered} \ \ \otimes\ \  \begin{gathered} \includegraphics[width=0.2\textwidth]{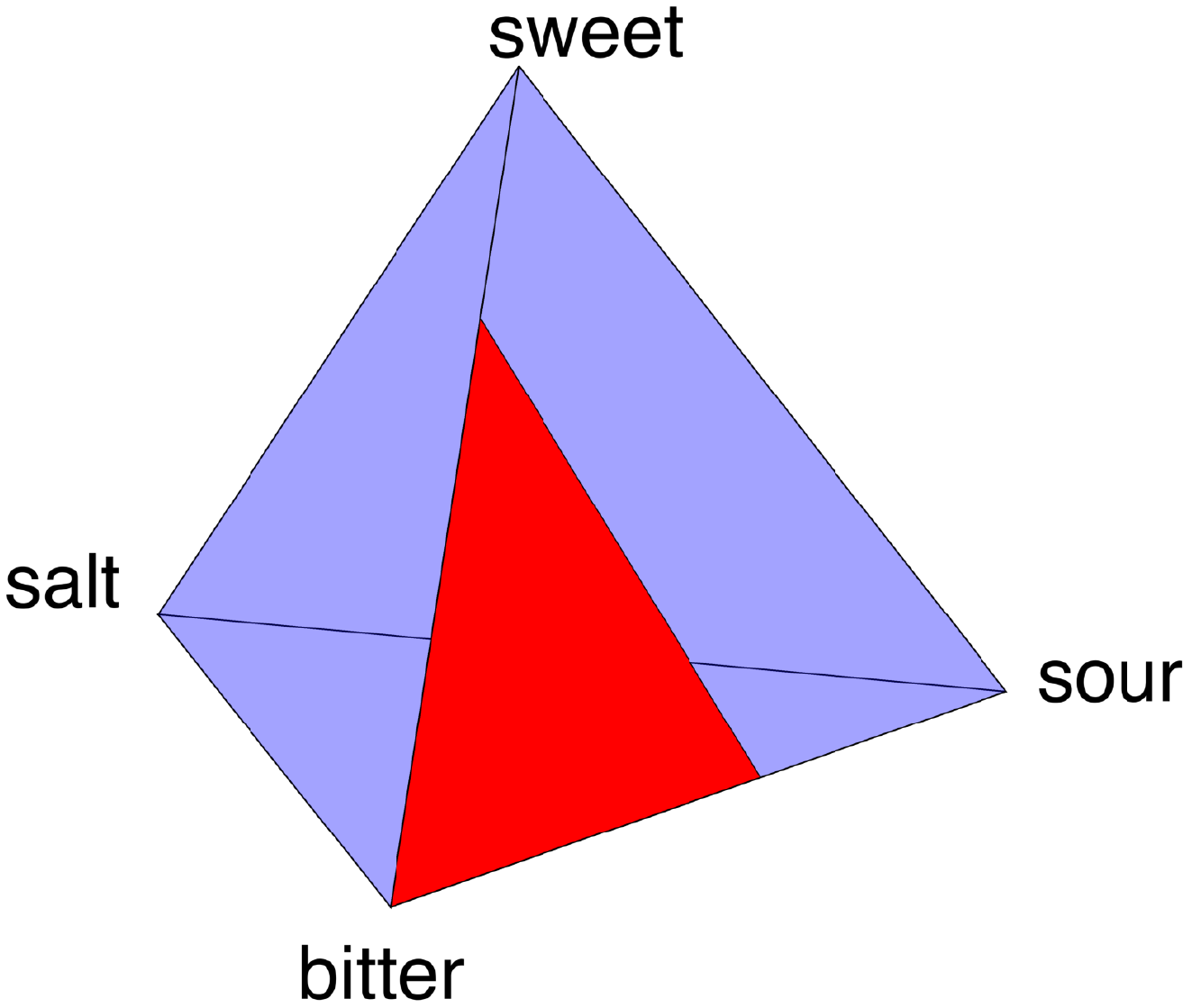}\end{gathered}\ \ \otimes\ \ \begin{gathered} \includegraphics[width=0.2\textwidth]{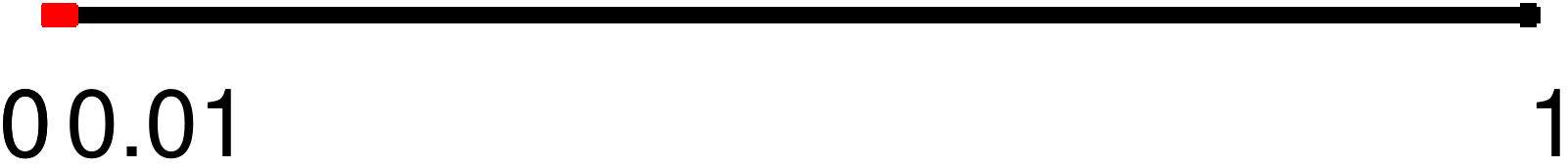}\end{gathered}\\
\end{align*}

What is an appropriate choice of sentence space for describing food and drink?
We need to describe the events associated with eating and drinking. We choose a very simple structure where the events are either positive or negative, and surprising or unsurprising. We therefore use a sentence space of pairs. The first element of the pair states whether the sentence is positive (1) or negative (0) and the second states whether it is surprising (1) or unsurprising (0). The convex structure on this space is the convex algebra on a join semilattice induced by element-wise max, as in example \ref{ex:semilattices}. We therefore have four points in the space: positive, surprising (1,~1); positive, unsurprising (1,~0); negative, surprising (0,~1); and negative, unsurprising (0,~0). Sentence meanings are convex subsets of this space, so they could be singletons, or larger subsets such as $\lang{negative} = \{(0,~1), (0,~0)\}$.

\subsubsection{Adjectives}
Recall that in a pregroup setting the adjective type is $n n^l$. In \convexrel, the adjective therefore has type $N\otimes N$.
Adjectives are convex relations on the noun space, so can be written as sets of ordered pairs. We give two examples, $\lang{yellow}_\lang{adj}$ and $\lang{soft}_\lang{adj}$.
The adjective~$\lang{yellow}_\lang{adj}$ has the simple form:
\[
\{(\vect{x}, \vect{x}) | x_\lang{colour} \in \property{yellow} \}
\]
This simple form reflects the fact that $\lang{yellow}_\lang{adj}$ depends only on one area of the conceptual space, so it really just corresponds to the property $\property{yellow}$.

An adjective such as `soft' behaves differently to this. We cannot simply define soft as one area of the conceptual space, because whether or not something is soft depends what it was originally. Using relations, we can start to write down the right type of structure for the adjective, as long as the objects are sufficiently distinct. Restricting our universe just to bananas and apples, we can write~$\lang{soft}_\lang{adj}$
as
\begin{equation*}
  \{(\vect{x}, \vect{x}) | \vect{x} \in \lang{banana} \text{ and } x_\lang{texture} \leq 0.35 \text{ or } \vect{x} \in \lang{apples} \text{ and } x_\lang{texture} \leq 0.6\}
\end{equation*} 
Note that here, we are using \lang{banana} and \lang{apple} as shorthand for specifications of convex areas of the conceptual space. These could be written out in longhand as sets of inequalities within the  colour and taste spaces.

An analysis of the difficulties in dealing with adjectives set-theoretically, breaking them down into (roughly) three categories, is given in~\cite{KampPartee1995}. 
Under this view, both adjectives and nouns are viewed as one-place predicates, so that, for example~$\lang{red} = \{ x | \text{$x$ is red}\}$ and~$\lang{dog} = \{ x |\text{$x$ is a dog}\}$.
There are then three classes of adjective. For \define{intersective} adjectives,  the meaning of $\lang{adj noun}$ is given by  $\lang{adj} \cap \lang{noun}$. For \define{subsective} adjectives, the meaning of $\lang{adj noun}$ is a subset of  $\lang{noun}$. For \define{privative} adjectives, however, $\lang{adj noun} \not\subseteq \lang{noun}$. 

Intersective adjectives are a simple modifier that can be thought of as the intersection between two concepts. 
We can make explicit the internal structure of these adjectives exploiting the multi-wires of theorem~\ref{thm:convexrel}.
For example, in the case of $\lang{yellow banana}$, we take the intersection of $\lang{yellow}$ and $\lang{banana}$.
We then show how to understand~$\lang{yellow}$ as an adjective. While the general case of  adjectives is depicted as: 
\[
\begin{gathered}\input{ftall.tikz}\end{gathered}\ \  =\ \  \begin{gathered}\input{tall_bent.tikz}\end{gathered}
\]
in the case of intersective adjectives the diagrams specialise to:
\begin{eqnarray*}
\begin{gathered}\input{yellow_int.tikz}\end{gathered}\ \ &=&\ \ \begin{gathered}\input{yellow_adj.tikz}\end{gathered}\\ \ \ &=&\ \  \begin{gathered}\input{yellow_bent.tikz}\end{gathered} 
\end{eqnarray*}
This shows us how the internal structure of an intersective adjective is derived directly from a noun.

\subsubsection{Verbs}
The pregroup type for a transitive verb is $n^r s n^l$, mapping to $N \otimes S\otimes N$ in \convexrel. To define the verb, we use concept names as shorthand, where these can easily be calculated. For example, is \lang{green} is considered to be an intersective adjective, \lang{green banana} can be calculated by taking the intersection of \lang{green} and \lang{banana} by combining the inequalities specifying the colour property, giving:
\begin{align*}
\lang{green banana} &= \{(R,G, B) | (R \leq G\leq 1.5R), (G \geq B), (0.3\leq R \leq 0.7), (G \geq 0.3)\}\\ 
&\qquad \otimes \ch{\{t_\lang{sweet}, 0.25t_\lang{sweet}+0.75t_\lang{bitter}, 0.7t_\lang{sweet}+0.3t_\lang{sour}\}} \otimes [0.2, 0.5]\\
\end{align*}

Although a full specification of a verb would take in all the nouns it could possibly apply to, for expository purposes we restrict our nouns to just bananas and beer which do not overlap, due to the fact that they have different textures. We define the verb ${\lang{taste} : I \rightarrow N \otimes S \otimes N}$ as follows:
\begin{align*}
\lang{taste} &= (\lang{green banana} \otimes \{(0, 0)\} \otimes \lang{bitter}) \cup  (\lang{green banana} \otimes \{(1, 1)\} \otimes \lang{sweet})\\
&\qquad  \cup (\lang{yellow banana} \otimes \{(1, 0)\} \otimes \lang{sweet})  \cup (\lang{beer} \otimes \{(0, 1)\}  \otimes \lang{sweet})\\ 
&\qquad \cup (\lang{beer} \otimes \{(1, 0)\} \otimes \lang{bitter})
\end{align*}

\subsection{Example: Robot movement}
We now present another example describing a simple formulation of robot movement. We will describe our choices of noun space~$N$ and sentence space~$S$, and show how to form nouns and verbs.

\subsubsection{Nouns}
The types of nouns we wish to describe are objects, such as \lang{armchair} and \lang{ball}, the robots \lang{Cathy} and \lang{David}, and places such as \lang{kitchen} and \lang{living room}. For shorthand, we call these nouns $a$, $b$, $c$, $d$, $k$, and $l$. These are specified in the noun space $N$ which is itself composed of a number of domains 
\[
N_\lang{location} \otimes N_\lang{direction} \otimes N_\lang{shape} \otimes N_\lang{size} \otimes N_\lang{colour}\otimes ...
\]

We firstly consider the kitchen and living rooms as being defined by convex subsets of points in the domain $N_\lang{location}$, defining properties in the location domain as:
\begin{align*}
p_\lang{kitchen location} &= \{(x_1, x_2) | x_1 \in [0, 5], x_2 \in [0, 10]\}\\
p_\lang{living room location} &= \{(x_1, x_2)| x_1 \in [5, 10], x_2 \in [0, 10]\}
\end{align*}
which can be depicted as follows:
\begin{center}
\input{rooms.tikz}
\end{center}
Then the nouns \lang{kitchen} and \lang{living room} are given by these properties together with other sets of characteristics in the shape domain, size domain, and so on, which we won't specify here.
\begin{align*}
\lang{kitchen} &= p_\lang{kitchen location} \otimes p_\lang{kitchen shape} \otimes p_\lang{kitchen size} \otimes ...\\
\lang{living room} &= p_\lang{living room location} \otimes p_\lang{living room shape} \otimes p_\lang{living room size} \otimes ...
\end{align*}
Similarly, the other nouns are defined by combinations of properties in the noun space. For this example, we do not worry too much about what they are, but assume that they allow us to distinguish between the objects.

\subsubsection{Verbs}
In order to define some verbs, we need consider what a suitable sentence space should look like. We want to give sentences of the form:
\begin{center}
 The ball is in the living room\\
Cathy moves to the kitchen
\end{center}
In these sentences, an object or agent is related to a path through time and space. Note that in the case of the verb `is in', this path is in fact trivially just a point, however for `moves to', the path actually is a path through time and space, and we will need to use subsets of the time and location domains to specify one single event. 
We therefore define the sentence space to be comprised of the noun space $N$, a time dimension $T$, and the location domain $N_\lang{location}$:
\[
S = N \otimes T \otimes N_\lang{location}
\]
The agent is represented by a point in the noun space $N$, and the path they take as described in the sentence is represented as a subset of the time and location domains. In what follows, we think of 0 on the time dimension $T$ as referring to `now', with negative values of $T$ referring to the past and positive values referring to the future.

As in the food example, transitive verbs are of the form $N\otimes S \otimes N$. This means that in this example, they are of the form 
\[
N\otimes N \otimes T \otimes N_\lang{location} \otimes N, 
\]
and can be thought of as sets of ordered tuples of the form 
\[
(n_1, n_2, t, l, n_3), 
\]
where $n_i$ stands for points in the noun space, $t$ is a time, $l$ is a location. We will consider the following verbs: \lang{is\_in}, \lang{moves\_to} \footnote{It could be argued that these are not transitive verbs, but intransitive verbs plus preposition. However, we can parse the combination as a transitive verb, since a preposition has type $s^r s n^l$ and therefore the combination reduces to type of a transitive verb:
\[
(n^r s)(s^r s n^l) \leq n^r s n^l
\]}.
The verb \lang{is\_in} can take any of the nouns $a$, $b$, $c$, or $d$ as subject, and any of $k$, $l$ as object. This verb refers to just one timepoint, i.e. now, or 0. The verb is as follows:
\begin{equation}
\label{eq:is}
\lang{is\_in} = \{(\vec{n}, \vec{n}, t_\lang{now}, m_\lang{location}, \vec{m})| \vec{n} \in a\cup b \cup c \cup d, t_\lang{now} = 0, \vec{m} \in k\cup l\}
\end{equation}
The verb \lang{moves\_to} refers to more than one point in time. We need to talk about an object moving from being at one location at a past time, to another location at time 0, or now. This movement should be continuous, since the objects we are talking about do not teleport from one point to another. We will also restrict the subject of the sentence to being one of the nouns $a$, $b$, $c$, or $d$, as we don't want to talk about the kitchen and living rooms moving at this point. The object of the verb, however, can be any of the nouns, so we can say, for example, that `Cathy moves to the armchair', or `The ball moves to Dave' (presumably because Cathy kicked it). The most specific event that can be described in the space will track the exact path that an object takes through space and time. The meaning of a less specific sentence will be a convex subset of these trajectories. We now define the verb as follows:  
\begin{align}
  \lang{moves\_to} = \{&(\vec{n}, \vec{n}, [t,0], f([t, 0]), \vec{m}) \mid \nonumber \\
   &\vec{n} \in a\cup b \cup c \cup d, t < 0,   f \text{ continuous}, f(t) \in \vec{n}_\lang{location}, f(0) \in \vec{m}_\lang{location}\} \label{eq:moves}
\end{align}
The constraints on~$t$ ensure that the movement happened in the past, and the constraints on $f$ ensure that the movement is from the location of the subject to the location of the object of the verb. These definitions are a little complex, but we will see how they work in interaction. Note that the location nouns \lang{kitchen} and \lang{living room} might seem to be of a different type to the object and agent nouns \lang{armchair}, \lang{ball}, \lang{Cathy} and \lang{David}. For example, we have specified that \lang{kitchen} and \lang{living room} do not move around. In future research we will be extending to a richer type system which can take account of these sorts of differences, and which will in fact be closer to that proposed by G\"ardenfors in \citep{Gardenfors2014}.

\section{Concepts in interaction}
\label{sec:composing}
We have given descriptions of how to form the different word types within our model of categorical conceptual spaces. In this section we show how to apply the type reductions of the pregroup grammar within the conceptual spaces formalism.

\subsection{Sentences in the food space}
The application of $\lang{yellow}_\lang{adj}$ to \lang{banana} works as follows.
\begin{align*}
\lang{yellow banana}
&= (1_N \otimes \epsilon_N)(\lang{yellow}_\lang{adj} \otimes \lang{banana})\\
&=(1_N \otimes \epsilon_N)\{(\vect{x}, \vect{x}) | x_\lang{colour} \in \lang{yellow} \} \\
 &\qquad\otimes (\{(R, G, B)|(0.9 R \leq G \leq 1.5 R), (R \geq 0.3), (B\leq 0.1)\} \\
&\qquad \otimes \ch{\{t_\lang{sweet}, 0.25t_\lang{sweet}+0.75t_\lang{bitter}, 0.7t_\lang{sweet}+0.3t_\lang{sour}\}} \otimes [0.2, 0.5])\\
&= \{(R, G, B)|(0.9 R \leq G \leq 1.5 R), (R \geq 0.7), (G \geq 0.7), (B\leq 0.1)\}\\
& \qquad \otimes \ch{\{t_\lang{sweet}, 0.25t_\lang{sweet}+0.75t_\lang{bitter}, 0.7t_\lang{sweet}+0.3t_\lang{sour}\}} \otimes [0.2, 0.5]
\end{align*}
Notice, in the last line, how the colour property has altered. This alteration restricts to yellow hues.
This assumes that we can tell bananas and apples apart by shape, colour and so on. Then the same calculation gives us 
\begin{align*}
\lang{soft apple} &=  \{(R, G, B)|R - 0.7 \leq G \leq R + 0.7), (G \geq 1-R), (B\leq 0.1)\} \\
&\qquad \otimes \ch{\{t_\lang{sweet}, 0.75t_\lang{sweet}+0.25t_\lang{bitter}, 0.3t_\lang{sweet}+0.7t_\lang{sour}\}} \otimes [0.4, 0.6]
\end{align*}

Using the definition of \lang{taste} that we gave, we find that although sweet bananas are good:
\begin{align*}
&\lang{bananas taste sweet} = (\epsilon_N \otimes 1_S \otimes \epsilon_N)(\lang{bananas}\otimes\lang{taste}\otimes\lang{sweet})\\
&\qquad = (\epsilon_N \otimes 1_S)(\lang{banana} \otimes (\lang{green banana} \otimes \{(1, 1)\} \cup \lang{yellow banana} \otimes \{(1, 0)\})\\\nonumber
&\qquad=\{(1,1), (1, 0)\} \\
&\qquad= \lang{positive}
\end{align*}
sweet beer is not so desirable:
\begin{align*}
\lang{beer tastes sweet} &= (\epsilon_N \otimes 1_S \otimes \epsilon_N)(\lang{beer}\otimes\lang{taste}\otimes\lang{sweet}) \\
&= \{(0, 1)\} \\
& = \lang{negative and surprising}
\end{align*}
\paragraph{Relative pronouns}
The compositional semantics we use can also deal with relative pronouns, described in detail in \cite{KartsaklisSadrzadehPulmanCoecke2013}. Relative pronouns are words such as `which'. For example, we can form the noun phrase \lang{Fruit which tastes bitter}. This has the  following structure:
\[
\begin{gathered}
\input{frob-sub-copy.tikz}
\end{gathered}
\]
which simplifies to: 
\[
\begin{gathered}
\input{frob-sub-copy-copy.tikz}
\end{gathered}
\]
In our example, we find that $\lang{Fruit which tastes bitter} = \lang{green banana}$:
\begin{align*}
\lang{Fruit which tastes bitter} 
&= (\mu_N \otimes \iota_S \otimes \epsilon_N)(\ch{\lang{bananas} \cup \lang{apples}} \otimes \lang{taste} \otimes \lang{bitter})\\
&=(\mu_N\otimes \iota_S)(\ch{\lang{bananas} \cup \lang{apples}} \otimes (\lang{green banana} \otimes \{(0, 0)\}))\\
&=\mu_N(\ch{\lang{bananas} \cup \lang{apples}} \otimes (\lang{green banana})) \\
&= \lang{green banana}
\end{align*}
where $\mu_N$ is the Frobenius merge map on $N$ and $\iota_S$ is the delete map on~$S$ described in theorem~\ref{thm:convexrel} and remark~\ref{rem:mergedel}.

\subsection{Sentences about robot movement}
In this section we describe how to compute the meaning of sentences about robot movement. 
Our first example is the sentence `Cathy moves to the living room'. In order to compute the meaning of this sentence, we assume that Cathy has a location. 
\begin{align}
&\lang{Cathy moves to the living room} \nonumber \\
&= (\epsilon_N \otimes 1_s \otimes \epsilon_N)(C \otimes \lang{moves\_to} \otimes L) \nonumber \\
&  = (\epsilon_N \otimes 1_s \otimes)(C \otimes \{(\vec{n},\vec{n}, [t, 0], f([t, 0]))|f(0) \in L_\lang{location}\}\label{eq:moreconstraints}\\
& = \{C, [t, 0], f([t, 0])|f(t) \in C_\lang{location}, f(0) \in L_\lang{location}\} \nonumber
\end{align}
In line~\eqref{eq:moreconstraints} further constraints apply to~$t$ and~$f$ as described in equation~\eqref{eq:moves}.
This calculation gives us a set of continuous line segments starting from Cathy's location at time $t$ and ending in the living room at time 0.

We now need to check that this set of line segments is convex. We assume that Cathy can take any possible location, and her other attributes remain static. This means that the set of possible instantiations of Cathy is convex. The set of time segments $[t, 0]$ such that $t < 0$ forms a convex set. Consider two such time segments. We define a convex mixture of these segments pointwise:
\[
p[t_1, 0] + (1 - p)[t_2, 0] = [pt_1 + (1 - p)t_2, p0 + (1 - p)0]
=[pt_1 + (1 - p)t_2, 0]
\] 
which clearly satisfies the condition that the start point is in the past and the end point is now.

We then consider the convex mixture of two sets of locations $f_1([t_1, 0])$ and $f_2([t_2, 0])$. In order to carry this out, we first of all transform the intervals $[t_1, 0]$ and $[t_2, 0]$ to $[-1, 0]$ by dividing through by $-t_i$, renaming the rescaled functions $f_i^\prime$. We then form a convex combination: 
\[
pf_1^\prime + (1-p) f_2^\prime : [0,1] \rightarrow N_\lang{location}  
\]
pointwise by taking:  
\[
(pf_1^\prime + (1-p) f_2^\prime)(\tau) = pf_1^\prime(\tau) + (1-p) f_2^\prime(\tau) = pf_1((-t_1)\tau) + (1-p) f_2((-t_2)\tau)
\]
Since both $f_1$ and $f_2$ are continuous in $T$, the result will be continuous in $T$.

The constraints on these sets are that $f_1(t_1)$ and $f_2(t_2)$ are in $C_\lang{location}$ and that $f_1(0) $ and $f_2(0)$ are in $L_\lang{location}$, and we need that their convex combinations are also in these respective locations. We know that $C_\lang{location}$ and $L_\lang{location}$ are convex, meaning that 
\[
pf_1(t_1) + (1-p)f_2(t_2) \in C_\lang{location}\qquad\quad\mbox{and}\qquad \quad pf_1(0) + (1-p)f_2(0) \in L_\lang{location}
\]
as required.

In this section, we have shown how we can use the interaction of words, represented by convex sets in conceptual spaces, to map sentence meanings down to a convex set in a conceptual space for sentences. 

\section{Discussion and related work}
In this paper we have shown how grammar can be introduced to conceptual spaces theory by extending the categorical compositional scheme of \cite{CoeckeSadrzadehClark2010}. By using the kind of transformational compositionality that linguistic grammar introduces, we are able to extend compositional accounts beyond simply using variants of conjunction or disjunction that are seen in \cite{aerts2009, hampton1987, hampton1988con, hampton1988dis, LewisLawry2016}. This means that adjective-noun, verb-noun, and full sentence composition may be modelled, and the meanings of phrases and sentences are mapped into a shared meaning space.

In the current work, we have restricted ourselves to grammatical composition, and in particular pregroup grammar. However, the categorical compositional scheme can be instantiated in a number of ways. The grammar can be changed from pregroup grammar to another categorial grammar, as in \cite{CoeckeGrefenstetteSadrzadeh2013}, or a compositional scheme that is not grammatically based may be used. Indeed, one of the challenges of this approach is to find a model of composition that accurately reflects human behaviour. One way of doing so would be to use an approach in which the syntactic scheme is generated by the semantics of the universe of discourse. Furthermore, since phrases and sentences are represented as sets equipped with a convex algebra, the model can in future work be extended to include logical composition.

\subsection{Related work}

Our general aim of integrating compositional and semantic aspects of cognition is a longstanding problem in AI. In the Integrated Connectionist-Symbolic framework \citep{smolensky2006}, the authors aim to integrate symbolic and connectionist reasoning, showing how symbolic reasoning can be instantiated in a neural network. One of the drawbacks of this approach is that sentence and phrase representations are dependent on the number of words in the phrase, so that it is difficult to directly compare noun phrases of different length, for example. That work inspired the development of the categorical compositional distributional semantic model of sentence meaning \citep{CoeckeSadrzadehClark2010} which the current work extends.

 Our research also has links to cognitive architectures which integrate compositional and semantic aspects of cognition. Examples of these are Nengo \citep{eliasmith2013}, neural blackboard architectures \citep{vandervelde2006}, and the previously mentioned \cite{smolensky2006}. Further, the use of conceptual spaces in cognitive architectures is an area of active research, as seen in, for example, \cite{forth2016,lieto2017}. Whilst the current work does not model key aspects of human cognition such as dynamics or action selection, it provides a model of compositionality that can be utilized by these architectures in representing inputs and how they may combine to form novel representations.

There is a wide range of research into modelling meaning and compositionality within conceptual spaces to which the theory developed here relates. Conceptual spaces have been formalized in \cite{RickardAisbettGibbon2007, AdamsRaubal2009, lawry2009}, and recently in \cite{bechberger2017}. However the type of compositionality defined covers conjunction, disjunction, and correlations between domains, whereas the framework we propose covers a far more general type of compositionality. In \cite{warglien2012a}, the authors develop a theory of verbs  within the conceptual spaces framework. That work presents a model of \emph{events} as consisting of at least a force vector and a result vector, together with a patient. It is also possible that an event has an agent, and an intention vector, amongst others. Verbs are then seen as being vectors that effect change, and which correspond only to one domain in the conceptual space. Our model has a similar goal in that it views a verb as a relation from one or more nouns to the sentence space. It therefore transforms nouns into sentences, which could be seen as events. However, rather than use a vector model of the verb, we use convex relations. \cite{derrac2015} use a data-driven approach to developing semantic relations within conceptual spaces. The relations obtained are modelled as directions within the conceptual space. \cite{mcgregor2016} develop a model of analogy within conceptual spaces, leveraging the geometrical properties of the meaning space.

\section{Conclusion and future work}
We have applied the categorical compositional scheme
to cognition and conceptual spaces. In order to do this we introduced a new model for categorical
compositional semantics, the category~\convexrel of convex algebras and binary relations respecting
convex structure. We consider this model as a proof of concept that we can describe convex structures within
our framework. Conceptual spaces are often considered to have further mathematical structure such as distance
measures and notions of convergence or fixed points. It is also possible to vary the notion of convexity under
consideration, for example by considering a binary betweenness relation on a space as primitive, rather than
a mixing operation. 

On the theoretical side, identifying a good setting for rich conceptual spaces models, and incorporating those
structures into a compositional framework is a direction for further work, building on \cite{MarsdenGenovese2017, CoeckeGenoveseLewisMarsdenToumi2017}.
 Other theoretical work to be done includes investigating the implementation of logic within \convexrel, specifically some form of negation, which in general does not preserve convexity.

On the more practical side, future work includes implementation within a data-derived conceptual space. This could include linguistic data derived from a corpus, but the generality of the conceptual spaces framework may also encompass a wider range of inputs such as from visual stimuli. Following on from this, future work will investigate integration with a cognitive architecture that encompasses dynamics and action.


\subsubsection*{Acknowledgements}
This work was partially funded by AFSOR grant ``Algorithmic and Logical Aspects when Composing Meanings'', the FQXi grant ``Categorical Compositional Physics'', and  EPSRC PhD scholarships.

\bibliographystyle{apalike}
\bibliography{cs}
\end{document}

%% file: tikzpreamble
\usepackage{tikz}
\usepackage{pgfplots}
\usetikzlibrary{trees}
\usetikzlibrary{topaths}
\usetikzlibrary{decorations.pathmorphing}
\usetikzlibrary{decorations.markings}
\usetikzlibrary{matrix,backgrounds,folding}
\usetikzlibrary{chains,scopes,positioning,fit}
\usetikzlibrary{arrows,shadows}
\usetikzlibrary{calc} 
\usetikzlibrary{chains}
\usetikzlibrary{shapes,shapes.geometric,shapes.misc}

\pgfdeclarelayer{edgelayer}
\pgfdeclarelayer{nodelayer}
\pgfsetlayers{background,edgelayer,nodelayer,main}

\tikzstyle{dot}=[circle,fill=black,draw=black]
\tikzstyle{none}=[inner sep=0pt]
\tikzstyle{every loop}=[]

\tikzstyle{(null)}=[]
\tikzstyle{plain}=[]

\tikzstyle{blank}=[inner sep=0pt]
\tikzstyle{box}=[rectangle, minimum size = 0.5cm,fill=white,draw=black]
\tikzstyle{small_node}=[inner sep=0pt, minimum size=0.2cm,circle,fill=white,draw=black]
\tikzstyle{square}=[thick, minimum size=0.3cm,rectangle,fill=white,draw=black]

\tikzstyle{downtri}=[regular polygon,regular polygon sides=3,shape border rotate=180,fill=white,draw=black]
\tikzstyle{uptri}=[regular polygon,regular polygon sides=3,shape border rotate=0,fill=white,draw=black]

\tikzstyle{morph}=[->,draw=black,line width=0.600]
\tikzstyle{arrow}=[-,draw=black,postaction={decorate},decoration={markings,mark=at position .5 with {\arrow{>}}},line width=2.000]
\tikzstyle{tick}=[-,draw=black,postaction={decorate},decoration={markings,mark=at position .5 with {\draw (0,-0.1) -- (0,0.1);}},line width=2.000]

%% file: wire.tikz
\begin{tikzpicture}
	\begin{pgfonlayer}{nodelayer}
		\node [style=blank] (0) at (0, 1) {};
		\node [style=blank] (1) at (0, -0) {};
		\node [style=blank] (2) at (-0.75, 0.5) {$A$};
	\end{pgfonlayer}
	\begin{pgfonlayer}{edgelayer}
		\draw (0) to (1); 
	\end{pgfonlayer}
\end{tikzpicture}

%% file: morphism1.tikz
\begin{tikzpicture}
	\begin{pgfonlayer}{nodelayer}
		\node [style=box] (0) at (0, 0) {$f$};
		\node [style=blank] (1) at (0, -0.75) {};
		\node [style=blank] (2) at (0, 0.75) {};
		\node [style=blank] (3) at (-0.75, -0.75) {$B$};
		\node [style=blank] (4) at (-0.75, 0.75) {$A$};
	\end{pgfonlayer}
	\begin{pgfonlayer}{edgelayer}
		\draw (1) to (0);
		\draw (2) to (0);
	\end{pgfonlayer}
\end{tikzpicture}

%% file: morphism2.tikz
\begin{tikzpicture}
	\begin{pgfonlayer}{nodelayer}
		\node [style=box] (0) at (0, 1.25) {$f$};
		\node [style=box] (1) at (0, 0) {$g$};
		\node [style=blank] (2) at (0, -0.75) {};
		\node [style=blank] (3) at (0, 2) {};
		\node [style=blank] (4) at (-0.75, 2) {$A$};
		\node [style=blank] (5) at (-0.75, 0.75) {$B$};
		\node [style=blank] (6) at (-0.75, -0.5) {$C$};
	\end{pgfonlayer}
	\begin{pgfonlayer}{edgelayer}
		\draw (1) to (2);
		\draw (1) to (0);
		\draw (0) to (3);
	\end{pgfonlayer}
\end{tikzpicture}

%% file: parallel.tikz
\begin{tikzpicture}
	\begin{pgfonlayer}{nodelayer}
		\node [style=box] (0) at (-1, 0) {$h$};
		\node [style=box] (1) at (0, 0) {$k$};
		\node [style=blank] (2) at (-1, -0.75) {};
		\node [style=blank] (3) at (0, -0.75) {};
		\node [style=blank] (4) at (-1, 0.75) {};
		\node [style=blank] (5) at (0, 0.75) {};
		\node [style=blank] (6) at (-1.5, -0.5) {$B$};
		\node [style=blank] (7) at (0.5, -0.5) {$D$};
		\node [style=blank] (8) at (0.5, 0.75) {$C$};
		\node [style=blank] (9) at (-1.5, 0.75) {$A$};
	\end{pgfonlayer}
	\begin{pgfonlayer}{edgelayer}
		\draw (2) to (0);
		\draw (0) to (4);
		\draw (1) to (5);
		\draw (1) to (3);
	\end{pgfonlayer}
\end{tikzpicture}

%% file: effect.tikz
\begin{tikzpicture}
	\begin{pgfonlayer}{nodelayer}
		\node [style=uptri] (0) at (0, -0) {$u$};
		\node [style=blank] (1) at (0, -0.75) {}; 
	\end{pgfonlayer}
	\begin{pgfonlayer}{edgelayer}
		\draw (0) to (1);
	\end{pgfonlayer}
\end{tikzpicture}

%% file: effect_copy.tikz
\begin{tikzpicture}
	\begin{pgfonlayer}{nodelayer}
		\node [style=downtri] (0) at (0, -0.75) {$v$};
		\node [style=blank] (1) at (0, 0) {};
	\end{pgfonlayer}
	\begin{pgfonlayer}{edgelayer}
		\draw (0) to (1);
	\end{pgfonlayer}
\end{tikzpicture}

%% file: cupscaps.tikz
\begin{tikzpicture}
	\begin{pgfonlayer}{nodelayer}
		\node [style=blank] (0) at (0, 0) {};
		\node [style=blank] (1) at (1, 0) {};
		\node [style=blank] (2) at (-1, 0) {};
		\node [style=blank] (3) at (-2, 0) {};
		\node [style=blank] (4) at (2, 0) {};
		\node [style=blank] (5) at (3, 0) {};
		\node [style=blank] (6) at (4, 0) {};
		\node [style=blank] (7) at (5, 0) {};
		\node [style=blank, anchor=mid] (8) at (2, -0.25) {$A^r$};
		\node [style=blank, anchor=mid] (9) at (3, -0.25) {$A$};
		\node [style=blank, anchor=mid] (10) at (4, -0.25) {$A$};
		\node [style=blank, anchor=mid] (11) at (5, -0.25) {$A^l$};
		\node [style=blank, anchor=mid] (12) at (1, 0.25) {$A$};
		\node [style=blank, anchor=mid] (13) at (0, 0.25) {$A^l$};
		\node [style=blank, anchor=mid] (14) at (-1, 0.25) {$A^r$}; 
		\node [style=blank, anchor=mid] (15) at (-2, 0.25) {$A$};
		\node [style=blank, anchor=mid] (16) at (-1.5, -1) {$\epsilon^r$};
		\node [style=blank, anchor=mid] (17) at (0.5, -1) {$\epsilon^l$};
		\node [style=blank, anchor=mid] (18) at (2.5, 1) {$\eta^r$};
		\node [style=blank, anchor=mid] (19) at (4.5, 1) {$\eta^l$};
	\end{pgfonlayer}
	\begin{pgfonlayer}{edgelayer}
		\draw [>->, bend right=90, looseness=1.75] (3) to (2);
		\draw [<-<,bend right=90, looseness=1.75] (0) to (1);
		\draw [>->, bend left=90, looseness=1.75] (4) to (5);
		\draw [<-<, bend left=90, looseness=1.50] (6) to (7);
	\end{pgfonlayer}
\end{tikzpicture}

%% file: snaaaakes.tikz
\begin{tikzpicture}
	\begin{pgfonlayer}{nodelayer}
		\node [style=blank] (0) at (1.5, 1.5) {};
		\node [style=blank] (1) at (2.5, 1.5) {};
		\node [style=blank] (2) at (3.5, 1.5) {};
		\node [style=blank] (3) at (3.5, 0.75) {};
		\node [style=blank] (4) at (1.5, 2.25) {};
		\node [style=blank] (5) at (-2.5, 1.5) {=};
		\node [style=blank] (6) at (-3.5, 2.25) {};
		\node [style=blank] (7) at (-3.5, 1.5) {};
		\node [style=blank] (8) at (-4.5, 1.5) {};
		\node [style=blank] (9) at (-5.5, 1.5) {};
		\node [style=blank] (10) at (-5.5, 0.75) {};
		\node [style=blank] (11) at (4.5, 1.5) {=};
		\node [style=blank] (12) at (5.5, 0.75) {};
		\node [style=blank] (13) at (5.5, 2.25) {};
		\node [style=blank, anchor=mid] (14) at (-5.5, 2.75) {$A$};
		\node [style=blank, anchor=mid] (15) at (-4.5, 2.75) {$A^l$};
		\node [style=blank, anchor=mid] (16) at (-3.5, 2.75) {$A$};
		\node [style=blank, anchor=mid] (17) at (1.5, 2.75) {$A$};
		\node [style=blank, anchor=mid] (18) at (2.5, 2.75) {$A^r$};
		\node [style=blank, anchor=mid] (19) at (3.5, 2.75) {$A$};
		\node [style=blank, anchor=mid] (20) at (5.5, 2.75) {$A$};
		\node [style=blank] (21) at (-1.5, 2.25) {};
		\node [style=blank, anchor=mid] (22) at (-1.5, 2.75) {$A$};
		\node [style=blank] (23) at (-1.5, 0.75) {};
		\node [style=blank, anchor=mid] (24) at (5.5, -0.75) {$A^r$};
		\node [style=blank] (25) at (3.5, -1.75) {};
		\node [style=blank] (26) at (5.5, -1) {};
		\node [style=blank] (27) at (1.5, -2.5) {};
		\node [style=blank] (28) at (2.5, -1.75) {};
		\node [style=blank, anchor=mid] (29) at (1.5, -0.75) {$A^r$};
		\node [style=blank, anchor=mid] (30) at (2.5, -0.75) {$A$};
		\node [style=blank, anchor=mid] (31) at (3.5, -0.75) {$A^r$};
		\node [style=blank] (32) at (3.5, -1) {};
		\node [style=blank] (33) at (5.5, -2.5) {};
		\node [style=blank, anchor=mid] (34) at (4.5, -1.75) {=};
		\node [style=blank] (35) at (1.5, -1.75) {};
		\node [style=blank, anchor=mid] (36) at (-2.5, -1.75) {=};
		\node [style=blank] (37) at (-5.5, -1) {}; 
		\node [style=blank] (38) at (-1.5, -1) {};
		\node [style=blank] (39) at (-1.5, -2.5) {};
		\node [style=blank] (40) at (-3.5, -2.5) {};
		\node [style=blank] (41) at (-3.5, -1.75) {};
		\node [style=blank, anchor=mid] (42) at (-4.5, -0.75) {$A$};
		\node [style=blank] (43) at (-5.5, -1.75) {};
		\node [style=blank, anchor=mid] (44) at (-3.5, -0.75) {$A^l$};
		\node [style=blank, anchor=mid] (45) at (-1.5, -0.75) {$A^l$};
		\node [style=blank, anchor=mid] (46) at (-5.5, -0.75) {$A^l$};
		\node [style=blank] (47) at (-4.5, -1.75) {};
	\end{pgfonlayer}
	\begin{pgfonlayer}{edgelayer}
		\draw [bend right=90, looseness=1.50] (0.center) to (1.center);
		\draw [bend left=90, looseness=1.50] (1.center) to (2.center);
		\draw [->](2.center) to (3.center);
		\draw [-<](0.center) to (4.center);
		\draw [bend left=90, looseness=1.50] (9.center) to (8.center);
		\draw [bend right=90, looseness=1.75] (8.center) to (7.center);
		\draw [-<](7.center) to (6.center);
		\draw [->](9.center) to (10.center);
		\draw [<-<](12.center) to (13.center);
		\draw [<-<](23.center) to (21.center);
		\draw [bend left=90, looseness=1.75] (35.center) to (28.center);
		\draw [bend right=90, looseness=1.75] (28.center) to (25.center);
		\draw [>-](27.center) to (35.center);
		\draw [->](25.center) to (32.center);
		\draw [>->](33.center) to (26.center);
		\draw [bend right=90, looseness=1.75] (43.center) to (47.center);
		\draw [bend left=90, looseness=1.50] (47.center) to (41.center);
		\draw [->](43.center) to (37.center);
		\draw [-<](41.center) to (40.center);
		\draw [>->](39.center) to (38.center);
	\end{pgfonlayer}
\end{tikzpicture}

%% file: chickens.tikz
\begin{tikzpicture}
	\begin{pgfonlayer}{nodelayer}
		\node [style=none] (0) at (0, -0) {};
		\node [style=none] (1) at (-0.25, -0) {};
		\node [style=none] (2) at (0.25, -0) {};
		\node [style=none] (3) at (1.75, -0) {};
		\node [style=none] (4) at (-1.75, -0) {};
		\node [style=none, anchor=mid] (5) at (-1.75, 0.5) {$n$};
		\node [style=none, anchor=mid] (6) at (0, 0.5) {$n^r s n^l $};
		\node [style=none, anchor=mid] (7) at (1.75, 0.5) {$n$};
		\node [style=none] (8) at (0, -1) {};
		\node [style=none, anchor=mid] (9) at (-1.75, 1.25) {\lang{chickens}};
		\node [style=none, anchor=mid] (10) at (0, 1.25) {\lang{cross}};
		\node [style=none, anchor=mid] (11) at (1.75, 1.25) {\lang{roads}};
	\end{pgfonlayer}
	\begin{pgfonlayer}{edgelayer}
		\draw (0.center) to (8.center);
		\draw [bend right=90, looseness=1.25] (4.center) to (1.center);
		\draw [bend right=90, looseness=1.25] (2.center) to (3.center);
	\end{pgfonlayer}
\end{tikzpicture}

%% file: chickens2.tikz
\begin{tikzpicture}
	\begin{pgfonlayer}{nodelayer}
		\node [style=none, anchor=mid] (0) at (-2.5, 1) {\lang{chickens}};
		\node [style=none, anchor=mid] (1) at (0, 1) {\lang{cross}};
		\node [style=none, anchor=mid] (2) at (2.5, 1) {\lang{roads}};
		\node [style=none] (3) at (-2.5, 0.75) {};
		\node [style=none] (4) at (-0.5, 0.75) {};
		\node [style=none] (5) at (0, 0.75) {};
		\node [style=none] (6) at (0.5, 0.75) {};
		\node [style=none] (7) at (2.5, 0.75) {};
		\node [style=none] (8) at (0, -0.25) {};
		\node [style=none] (9) at (3.75, 0.75) {};
		\node [style=none] (10) at (1.25, 0.75) {};
		\node [style=none] (11) at (2.5, 1.75) {};
		\node [style=none] (12) at (-2.5, 1.75) {};
		\node [style=none] (13) at (-3.75, 0.75) {};
		\node [style=none] (14) at (-1.25, 0.75) {};
		\node [style=none] (15) at (0, 1.75) {};
		\node [style=none] (16) at (-1, 0.75) {};
		\node [style=none] (17) at (1, 0.75) {};
		\node [style=none] (18) at (-1.5, 0) {$N$};
		\node [style=none] (19) at (1.5, 0) {$N$};
		\node [style=none] (20) at (0.25, 0) {$S$};
	\end{pgfonlayer}
	\begin{pgfonlayer}{edgelayer}
		\draw [bend right=90, looseness=0.75] (3.center) to (4.center);
		\draw [bend right=90, looseness=0.75] (6.center) to (7.center);
		\draw (5.center) to (8.center);
		\draw [style=swap] (11.center) to (9.center);
		\draw [style=swap] (9.center) to (10.center);
		\draw [style=swap] (10.center) to (11.center);
		\draw [style=swap] (12.center) to (14.center);
		\draw [style=swap] (14.center) to (13.center);
		\draw [style=swap] (13.center) to (12.center);
		\draw [style=swap] (15.center) to (17.center);
		\draw [style=swap] (17.center) to (16.center);
		\draw [style=swap] (16.center) to (15.center);
	\end{pgfonlayer}
\end{tikzpicture}

%% file: chickens3.tikz
\begin{tikzpicture}
	\begin{pgfonlayer}{nodelayer}
		\node [style=none, anchor=mid] (0) at (-5.5, 1) {\lang{chickens}};
		\node [style=none, anchor=mid] (1) at (-0.5, 1) {\lang{cross}};
		\node [style=none, anchor=mid] (2) at (2, 1) {\lang{roads}};
		\node [style=none] (3) at (-5.5, 0.75) {};
		\node [style=none] (4) at (-3.5, 0.75) {};
		\node [style=none] (5) at (0, 0.75) {};
		\node [style=none] (6) at (2, 0.75) {};
		\node [style=none] (7) at (3.25, 0.75) {};
		\node [style=none] (8) at (0.75, 0.75) {};
		\node [style=none] (9) at (2, 1.75) {};
		\node [style=none] (10) at (-5.5, 1.75) {};
		\node [style=none] (11) at (-6.75, 0.75) {};
		\node [style=none] (12) at (-4.25, 0.75) {};
		\node [style=none] (13) at (-0.5, 1.75) {};
		\node [style=none] (14) at (-1.5, 0.75) {};
		\node [style=none] (15) at (0.5, 0.75) {};
		\node [style=none] (16) at (-3, 0.75) {};
		\node [style=none] (17) at (-2.5, 0.75) {};
		\node [style=none] (18) at (-3.5, 0.75) {};
		\node [style=none] (19) at (-2, 0.75) {};
		\node [style=none] (20) at (-2.5, 0.75) {};
		\node [style=none] (21) at (-1, 0.75) {};
		\node [style=none] (22) at (-0.5, 0.75) {};
		\node [style=none] (23) at (-2, 0.75) {};
		\node [style=none] (24) at (-3, 0.75) {};
		\node [style=none] (25) at (-3, -0.25) {};
		\node [style=none, anchor=mid] (26) at (-2.75, 1.5) {\lang{do}};
		\node [style=none] (27) at (-2.75, 0) {$S$};
		\node [style=none] (28) at (-4.5, 0) {$N$};
		\node [style=none] (29) at (1, 0) {$N$};
	\end{pgfonlayer}
	\begin{pgfonlayer}{edgelayer}
		\draw [bend right=90, looseness=0.75] (3.center) to (4.center);
		\draw [bend right=90, looseness=0.75] (5.center) to (6.center);
		\draw [style=swap] (9.center) to (7.center);
		\draw [style=swap] (7.center) to (8.center);
		\draw [style=swap] (8.center) to (9.center);
		\draw [style=swap] (10.center) to (12.center);
		\draw [style=swap] (12.center) to (11.center);
		\draw [style=swap] (11.center) to (10.center);
		\draw [style=swap] (13.center) to (15.center);
		\draw [style=swap] (15.center) to (14.center);
		\draw [style=swap] (14.center) to (13.center);
		\draw [bend left=90, looseness=1.00] (16.center) to (17.center);
		\draw [bend left=90, looseness=1.00] (18.center) to (19.center);
		\draw [bend right=90, looseness=0.75] (23.center) to (21.center);
		\draw [bend right=90, looseness=0.75] (20.center) to (22.center);
		\draw (24.center) to (25.center);
	\end{pgfonlayer}
\end{tikzpicture}

%% file: spider2a.tikz
\begin{tikzpicture}[scale=0.7, text height=1.5 ex]
	\begin{pgfonlayer}{nodelayer}
		\node [style=none] (0) at (-0.75, 0.75) {};
		\node [style={small_node}] (1) at (0, 0) {};
		\node [style=none] (2) at (0.75, 0.75) {};
		\node [style=none] (3) at (0, -0.75) {};
	\end{pgfonlayer}
	\begin{pgfonlayer}{edgelayer}
		\draw [swap, in=180, out=-90, looseness=1.25] (0.center) to (1);
		\draw [swap] (1) to (3.center);
		\draw [swap, in=0, out=-90, looseness=1.25] (2.center) to (1);
	\end{pgfonlayer}
\end{tikzpicture}

%% file: spider2b.tikz
\begin{tikzpicture}[scale=0.7, text height=1.5 ex]
	\begin{pgfonlayer}{nodelayer}
		\node [style=none] (0) at (0, 0.5) {};
		\node [style={small_node}] (1) at (0, -0.25) {};
	\end{pgfonlayer}
	\begin{pgfonlayer}{edgelayer}
		\draw [swap] (1) to (0.center);
	\end{pgfonlayer}
\end{tikzpicture}

%% file: spider1.tikz
\begin{tikzpicture}[scale=0.7, text height=1.5 ex]
	\begin{pgfonlayer}{nodelayer}
		\node [style=none] (0) at (-3.5, 1.75) {};
		\node [style={small_node}] (1) at (-2.75, 0.75) {};
		\node [style=none] (2) at (-2, 1.75) {};
		\node [style=none] (3) at (-2, 0) {};
		\node [style={small_node}] (4) at (-2.75, 0.75) {};
		\node [style=none] (5) at (-3.5, 0) {};
		\node [style=none] (6) at (-0.5, 0) {};
		\node [style=none] (7) at (-2, -1.75) {};
		\node [style=none] (8) at (-0.5, -1.75) {};
		\node [style={small_node}] (9) at (-1.25, -0.75) {};
		\node [style=none] (10) at (-2, 0) {};
		\node [style={small_node}] (11) at (-1.25, -0.75) {};
		\node [style=none] (12) at (3.5, 1) {};
		\node [style=none] (13) at (2, -1) {};
		\node [style=none] (14) at (3.5, -1) {};
		\node [style={small_node}] (15) at (2.75, 0) {};
		\node [style=none] (16) at (2, 1) {};
		\node [style={small_node}] (17) at (2.75, 0) {};
		\node [style=none] (18) at (0.75, 0) {$=$};
		\node [style=none] (19) at (-3.5, -1.75) {};
		\node [style=none] (20) at (-0.5, 1.75) {}; 
		\node [style=none] (21) at (2.75, 0.75) {$\ldots$};
		\node [style=none] (22) at (2.75, -0.75) {$\ldots$};
		\node [style=none] (23) at (-1.25, -1.5) {$\ldots$};
		\node [style=none] (24) at (-2.75, -1.5) {$\ldots$};
		\node [style=none] (25) at (-1.25, 1.5) {$\ldots$};
		\node [style=none] (26) at (-2.75, 1.5) {$\ldots$};
	\end{pgfonlayer}
	\begin{pgfonlayer}{edgelayer}
		\draw [swap, in=180, out=-90, looseness=1.00] (0.center) to (1);
		\draw [swap, in=15, out=-90, looseness=1.25] (2.center) to (1);
		\draw [swap, in=-165, out=90, looseness=1.25] (5.center) to (4);
		\draw [swap, in=-15, out=90, looseness=1.25] (3.center) to (4);
		\draw [swap, in=165, out=-90, looseness=1.25] (10.center) to (9);
		\draw [swap, in=15, out=-90, looseness=1.25] (6.center) to (9);
		\draw [swap, in=-165, out=90, looseness=1.25] (7.center) to (11);
		\draw [swap, in=-15, out=90, looseness=1.25] (8.center) to (11);
		\draw [swap, in=165, out=-90, looseness=1.25] (16.center) to (15);
		\draw [swap, in=15, out=-90, looseness=1.25] (12.center) to (15);
		\draw [swap, in=-165, out=90, looseness=1.25] (13.center) to (17);
		\draw [swap, in=-15, out=90, looseness=1.25] (14.center) to (17);
		\draw [style=swap] (20.center) to (6.center);
		\draw [style=swap] (5.center) to (19.center);
	\end{pgfonlayer}
\end{tikzpicture}

%% file: chickens4.tikz
\begin{tikzpicture}
	\begin{pgfonlayer}{nodelayer}
		\node [style=none, anchor=mid] (0) at (-5.5, 1) {\lang{chickens}};
		\node [style=none, anchor=mid] (1) at (-0.5, 1) {\lang{cross}};
		\node [style=none, anchor=mid] (2) at (2, 1) {\lang{roads}};
		\node [style=none] (3) at (-5.5, 0.75) {};
		\node [style=none] (4) at (-3.5, 0.75) {};
		\node [style=none] (5) at (0, 0.75) {};
		\node [style=none] (6) at (2, 0.75) {};
		\node [style=none] (7) at (3.25, 0.75) {};
		\node [style=none] (8) at (0.75, 0.75) {};
		\node [style=none] (9) at (2, 1.75) {};
		\node [style=none] (10) at (-5.5, 1.75) {};
		\node [style=none] (11) at (-6.75, 0.75) {};
		\node [style=none] (12) at (-4.25, 0.75) {};
		\node [style=none] (13) at (-0.5, 1.75) {};
		\node [style=none] (14) at (-1.5, 0.75) {};
		\node [style=none] (15) at (0.5, 0.75) {};
		\node [style=none] (16) at (-3.5, 0.75) {};
		\node [style=none] (17) at (-2, 0.75) {};
		\node [style=none] (18) at (-2.5, 0.75) {};
		\node [style=none] (19) at (-1, 0.75) {};
		\node [style=none] (20) at (-0.5, 0.75) {};
		\node [style=none] (21) at (-2, 0.75) {};
		\node [style=none] (22) at (-3, 1.25) {};
		\node [style=none] (23) at (-3, -0.25) {};
		\node [style=none, anchor=mid] (24) at (-2.75, 1.75) {\lang{that}};
		\node [style={small_node}] (25) at (-3, 1.25) {};
		\node [style={small_node}] (26) at (-2.5, 0.75) {};
		\node [style=none] (27) at (-4.5, 0) {$N$};
		\node [style=none] (28) at (1, 0) {$N$};
		\node [style=none] (29) at (-2.75, 0) {$N$};
		\node [style=none] (30) at (-1.5, 0) {$S$};
	\end{pgfonlayer}
	\begin{pgfonlayer}{edgelayer}
		\draw [bend right=90, looseness=0.75] (3.center) to (4.center);
		\draw [bend right=90, looseness=0.75] (5.center) to (6.center);
		\draw [style=swap] (9.center) to (7.center);
		\draw [style=swap] (7.center) to (8.center);
		\draw [style=swap] (8.center) to (9.center);
		\draw [style=swap] (10.center) to (12.center);
		\draw [style=swap] (12.center) to (11.center);
		\draw [style=swap] (11.center) to (10.center);
		\draw [style=swap] (13.center) to (15.center);
		\draw [style=swap] (15.center) to (14.center);
		\draw [style=swap] (14.center) to (13.center);
		\draw [bend right=90, looseness=0.75] (21.center) to (19.center);
		\draw [in=-90, out=-90, looseness=0.75] (18.center) to (20.center);
		\draw (22.center) to (23.center);
		\draw [style=swap, in=180, out=90, looseness=1.00] (16.center) to (22.center);
		\draw [style=swap, in=90, out=0, looseness=1.00] (22.center) to (21.center);
	\end{pgfonlayer}
\end{tikzpicture}

%% file: cap.tikz
\begin{tikzpicture}
	\begin{pgfonlayer}{nodelayer}
		\node [style=blank] (0) at (-0.5, -0.25) {};
		\node [style=blank] (1) at (0.5, -0.25) {};
	\end{pgfonlayer}
	\begin{pgfonlayer}{edgelayer}
		\draw [bend left=90, looseness=1.50] (0) to (1);
	\end{pgfonlayer}
\end{tikzpicture}

%% file: cup2.tikz
\begin{tikzpicture}
	\begin{pgfonlayer}{nodelayer}
		\node [style=blank] (0) at (-0.5, -0.25) {};
		\node [style=blank] (1) at (0.5, -0.25) {};
	\end{pgfonlayer}
	\begin{pgfonlayer}{edgelayer}
		\draw [bend right=90, looseness=1.50] (0) to (1);
	\end{pgfonlayer}
\end{tikzpicture}

%% file: spider.tikz
\begin{tikzpicture}[scale=0.7, text height=1.5 ex]
	\begin{pgfonlayer}{nodelayer}
		\node [style=none] (0) at (0.75, 1) {};
		\node [style=none] (1) at (-0.75, -1) {};
		\node [style=none] (2) at (0.75, -1) {};
		\node [style={small_node}] (3) at (0, 0) {};
		\node [style=none] (4) at (-0.75, 1) {};
		\node [style={small_node}] (5) at (0, 0) {};
		\node [style=none] (6) at (0, 0.75) {$\ldots$};
		\node [style=none] (7) at (0, -0.75) {$\ldots$};
	\end{pgfonlayer}
	\begin{pgfonlayer}{edgelayer}
		\draw [swap, in=165, out=-90, looseness=1.25] (4.center) to (3);
		\draw [swap, in=15, out=-90, looseness=1.25] (0.center) to (3);
		\draw [swap, in=-165, out=90, looseness=1.25] (1.center) to (5);
		\draw [swap, in=-15, out=90, looseness=1.25] (2.center) to (5);
	\end{pgfonlayer}
\end{tikzpicture}

%% file: ftall.tikz
\begin{tikzpicture}
	\begin{pgfonlayer}{nodelayer}
		\node [style=none] (0) at (0, 0.5) {};
		\node [style=none] (1) at (0, 0) {};
		\node [style=none] (2) at (0, -1) {};
		\node [style=none] (3) at (0, -1.75) {};
		\node [style=none] (4) at (-0.5, 0) {};
		\node [style=none] (5) at (-0.5, -1) {};
		\node [style=none] (6) at (0.5, -1) {};
		\node [style=none] (7) at (0.5, 0) {};
		\node [style=none, anchor=mid] (8) at (0, -0.5) {\lang{soft}};
		\node [style=none] (9) at (0.25, -1.5) {$N$};
		\node [style=none, anchor=mid] (10) at (0, 0.75) {\lang{banana}};
		\node [style=none] (11) at (0, 0.5) {};
		\node [style=none] (12) at (0, 1.5) {};
		\node [style=none] (13) at (-1.25, 0.5) {};
		\node [style=none] (14) at (1.25, 0.5) {};
		\node [style=none] (15) at (0.25, 0.25) {$N$};
	\end{pgfonlayer}
	\begin{pgfonlayer}{edgelayer}
		\draw (0.center) to (1.center);
		\draw (2.center) to (3.center);
		\draw (4.center) to (5.center);
		\draw (5.center) to (6.center);
		\draw (6.center) to (7.center);
		\draw (7.center) to (4.center);
		\draw [style=swap] (12.center) to (14.center);
		\draw [style=swap] (14.center) to (13.center);
		\draw [style=swap] (13.center) to (12.center);
	\end{pgfonlayer}
\end{tikzpicture}

%% file: tall_bent.tikz
\begin{tikzpicture}
	\begin{pgfonlayer}{nodelayer}
		\node [style=none] (0) at (1.25, -0.5) {};
		\node [style=none] (1) at (-1.5, 0.75) {};
		\node [style=none] (2) at (-1.5, -0.25) {};
		\node [style=none, anchor=mid] (3) at (-2, 0.25) {\lang{soft}};
		\node [style=none] (4) at (1.25, -0.5) {};
		\node [style=none] (5) at (-2.75, -0.5) {};
		\node [style=none] (6) at (-2.75, -1.5) {};
		\node [style=none] (7) at (-2.5, 0.75) {};
		\node [style=none] (8) at (-0.25, -0.5) {};
		\node [style=none] (9) at (-2, 1.5) {};
		\node [style=none] (10) at (-3.75, -0.5) {};
		\node [style=none] (11) at (-2.5, 0.25) {};
		\node [style=none] (12) at (-1.5, 0.25) {};
		\node [style=none] (13) at (-1.25, -0.5) {};
		\node [style=none] (14) at (-2.5, -0.25) {};
		\node [style=none] (15) at (0, -1.25) {$N$};
		\node [style=none] (16) at (-1.25, -0.5) {};
		\node [style=none] (17) at (1.25, 0.5) {};
		\node [style=none, anchor=mid] (18) at (1.25, -0.25) {\lang{banana}};
		\node [style=none] (19) at (2.5, -0.5) {};
		\node [style=none] (20) at (1.25, -0.5) {};
		\node [style=none] (21) at (0, -0.5) {};
		\node [style=none] (22) at (-2.5, -1.25) {$N$};
	\end{pgfonlayer}
	\begin{pgfonlayer}{edgelayer}
		\draw (5.center) to (6.center);
		\draw (1.center) to (2.center);
		\draw (7.center) to (1.center);
		\draw [in=0, out=90, looseness=1.00] (13.center) to (12.center);
		\draw [in=90, out=180, looseness=1.00] (11.center) to (5.center);
		\draw (8.center) to (9.center);
		\draw (9.center) to (10.center);
		\draw (8.center) to (10.center);
		\draw (14.center) to (2.center);
		\draw (14.center) to (7.center);
		\draw [bend right=90, looseness=0.75] (16.center) to (4.center);
		\draw [style=swap] (17.center) to (19.center);
		\draw [style=swap] (19.center) to (21.center);
		\draw [style=swap] (21.center) to (17.center);
	\end{pgfonlayer}
\end{tikzpicture}

%% file: yellow_int.tikz
\begin{tikzpicture}
	\begin{pgfonlayer}{nodelayer}
		\node [style=none] (0) at (-1.5, 0.5) {};
		\node [style=none, anchor=mid] (1) at (-1.5, 0.75) {\lang{yellow}};
		\node [style={small_node}] (2) at (0, 0) {};
		\node [style=none] (3) at (1.5, 0.5) {};
		\node [style=none] (4) at (0.25, -0.5) {$N$};
		\node [style=none] (5) at (0, -0.75) {};
		\node [style=none] (6) at (1.5, 0.5) {};
		\node [style=none] (7) at (2.75, 0.5) {};
		\node [style=none] (8) at (0.25, 0.5) {};
		\node [style=none, anchor=mid] (9) at (1.5, 0.75) {\lang{banana}};
		\node [style=none] (10) at (1.5, 0.5) {};
		\node [style=none] (11) at (1.5, 0.5) {};
		\node [style=none] (12) at (1.5, 1.5) {};
		\node [style=none] (13) at (-0.25, 0.5) {};
		\node [style=none] (14) at (-2.75, 0.5) {};
		\node [style=none] (15) at (-1.5, 1.5) {};
	\end{pgfonlayer}
	\begin{pgfonlayer}{edgelayer}
		\draw [in=180, out=-90, looseness=0.75] (0.center) to (2);
		\draw (2) to (5.center);
		\draw [in=0, out=-90, looseness=0.75] (3.center) to (2);
		\draw [style=swap] (12.center) to (7.center);
		\draw [style=swap] (7.center) to (8.center);
		\draw [style=swap] (8.center) to (12.center);
		\draw [style=swap] (15.center) to (13.center);
		\draw [style=swap] (13.center) to (14.center);
		\draw [style=swap] (14.center) to (15.center);
	\end{pgfonlayer}
\end{tikzpicture}

%% file: yellow_adj.tikz
\begin{tikzpicture}
	\begin{pgfonlayer}{nodelayer}
		\node [style=none] (0) at (0, 1) {};
		\node [style=none] (1) at (0, -1.5) {};
		\node [style=none] (2) at (0, -2.25) {};
		\node [style=none] (3) at (-1.75, 0.75) {};
		\node [style=none] (4) at (-1.75, -1.5) {};
		\node [style=none] (5) at (1.75, -1.5) {};
		\node [style=none] (6) at (1.75, 0.75) {};
		\node [style=none] (7) at (0, -0.75) {};
		\node [style={small_node}] (8) at (0, -1.25) {};
		\node [style=none] (9) at (1.5, -0.5) {};
		\node [style=none] (10) at (0, 0.75) {};
		\node [style=none, anchor=mid] (11) at (0, 1.25) {\lang{banana}};
		\node [style=none] (12) at (-1.25, 1) {};
		\node [style=none] (13) at (0, 2) {};
		\node [style=none] (14) at (1.25, 1) {};
		\node [style=none] (15) at (0, 0.25) {};
		\node [style=none] (16) at (-1.25, -0.75) {};
		\node [style=none] (17) at (1.25, -0.75) {};
		\node [style=none, anchor=mid] (18) at (0, -0.5) {\lang{yellow}};
		\node [style=none] (19) at (-0.5, -1) {};
		\node [style=none] (20) at (0.25, -2) {$N$};
	\end{pgfonlayer}
	\begin{pgfonlayer}{edgelayer}
		\draw (1.center) to (2.center);
		\draw (3.center) to (4.center);
		\draw (4.center) to (5.center);
		\draw (5.center) to (6.center);
		\draw (6.center) to (3.center);
		\draw (8) to (1.center);
		\draw [in=0, out=-90, looseness=0.75] (9.center) to (8);
		\draw (0.center) to (10.center);
		\draw [style=swap] (13.center) to (14.center);
		\draw [style=swap] (14.center) to (12.center);
		\draw [style=swap] (12.center) to (13.center);
		\draw [style=swap] (15.center) to (17.center);
		\draw [style=swap] (17.center) to (16.center);
		\draw [style=swap] (16.center) to (15.center);
		\draw [style=swap, in=-90, out=90, looseness=1.00] (19.center) to (7.center);
		\draw [style=swap, in=165, out=-90, looseness=0.75] (19.center) to (8);
		\draw [style=swap, in=90, out=-90, looseness=0.75] (10.center) to (9.center);
	\end{pgfonlayer}
\end{tikzpicture}

%% file: yellow_bent.tikz
\begin{tikzpicture}
	\begin{pgfonlayer}{nodelayer}
		\node [style=none] (0) at (0, 0) {};
		\node [style={small_node}] (1) at (0, -0.25) {};
		\node [style=none] (2) at (-0.5, -0.5) {};
		\node [style=none] (3) at (3.75, -0.5) {};
		\node [style=none] (4) at (2.25, -1.25) {$N$};
		\node [style=none] (5) at (0.5, -0.5) {};
		\node [style=none] (6) at (0, 1.25) {};
		\node [style=none] (7) at (-2.25, -0.5) {};
		\node [style=none] (8) at (2.25, -0.5) {};
		\node [style=none] (9) at (-0.5, -1.5) {};
		\node [style=none] (10) at (0.5, -0.5) {};
		\node [style=none] (11) at (5, -0.5) {};
		\node [style=none] (12) at (3.75, -0.5) {};
		\node [style=none, anchor=mid] (13) at (3.75, -0.25) {\lang{banana}};
		\node [style=none] (14) at (3.75, 0.5) {};
		\node [style=none] (15) at (2.5, -0.5) {};
		\node [style=none] (16) at (1.25, 0) {};
		\node [style=none] (17) at (0, 1) {};
		\node [style=none] (18) at (0, 0) {};
		\node [style=none, anchor=mid] (19) at (0, 0.25) {\lang{yellow}};
		\node [style=none] (20) at (-1.25, 0) {};
		\node [style=none] (21) at (-0.25, -1.25) {$N$};
	\end{pgfonlayer}
	\begin{pgfonlayer}{edgelayer}
		\draw [in=90, out=-90, looseness=1.00] (0.center) to (1);
		\draw [in=90, out=180, looseness=1.00] (1) to (2.center);
		\draw [in=90, out=0, looseness=1.00] (1) to (5.center);
		\draw (6.center) to (7.center);
		\draw (7.center) to (8.center);
		\draw (6.center) to (8.center);
		\draw (2.center) to (9.center);
		\draw [bend right=90, looseness=0.50] (10.center) to (3.center);
		\draw [style=swap] (14.center) to (11.center);
		\draw [style=swap] (11.center) to (15.center);
		\draw [style=swap] (15.center) to (14.center);
		\draw [style=swap] (17.center) to (16.center);
		\draw [style=swap] (16.center) to (20.center);
		\draw [style=swap] (20.center) to (17.center);
	\end{pgfonlayer}
\end{tikzpicture}

%% file: rooms.tikz
\begin{tikzpicture}
	\begin{pgfonlayer}{nodelayer}
		\node [style=none] (0) at (-2, 2) {};
		\node [style=none] (1) at (2, 2) {};
		\node [style=none] (2) at (2, -2) {};
		\node [style=none] (3) at (-2, -2) {};
		\node [style=none] (4) at (0, 2) {};
		\node [style=none] (5) at (0, -2) {};
		\node [style=none] (6) at (0, -2) {};
		\node [style=none, anchor=mid] (7) at (-1, 0) {$\lang{kitchen}$};
		\node [style=none, anchor=mid] (8) at (1, 0) {$\lang{living room}$};
		\node [style=none] (9) at (-2, -2.25) {0};
		\node [style=none] (10) at (0, -2.25) {5};
		\node [style=none] (11) at (2, -2.25) {10};
		\node [style=none] (12) at (-2.25, -2) {0};
		\node [style=none] (13) at (-2.25, 2) {10};
		\node [style=none] (14) at (0, -2.75) {$x_1$};
		\node [style=none] (15) at (-2.5, 0) {$x_2$};
	\end{pgfonlayer}
	\begin{pgfonlayer}{edgelayer}
		\draw (0.center) to (3.center);
		\draw (3.center) to (2.center);
		\draw (2.center) to (1.center);
		\draw (1.center) to (0.center);
		\draw (4.center) to (5.center);
	\end{pgfonlayer}
\end{tikzpicture}

%% file: frob-sub-copy.tikz
\begin{tikzpicture}
	\begin{pgfonlayer}{nodelayer}
		\node [style=none, anchor=mid] (0) at (-5.5, 1) {\lang{fruit}};
		\node [style=none, anchor=mid] (1) at (-0.5, 1) {\lang{tastes}};
		\node [style=none, anchor=mid] (2) at (2, 1) {\lang{bitter}};
		\node [style=none] (3) at (-5.5, 0.75) {};
		\node [style=none] (4) at (-3.5, 0.75) {};
		\node [style=none] (5) at (0, 0.75) {};
		\node [style=none] (6) at (2, 0.75) {};
		\node [style=none] (7) at (3.25, 0.75) {};
		\node [style=none] (8) at (0.75, 0.75) {};
		\node [style=none] (9) at (2, 1.75) {};
		\node [style=none] (10) at (-5.5, 1.75) {};
		\node [style=none] (11) at (-6.75, 0.75) {};
		\node [style=none] (12) at (-4.25, 0.75) {};
		\node [style=none] (13) at (-0.5, 1.75) {};
		\node [style=none] (14) at (-1.5, 0.75) {};
		\node [style=none] (15) at (0.5, 0.75) {};
		\node [style=none] (16) at (-3.5, 0.75) {};
		\node [style=none] (17) at (-2, 0.75) {};
		\node [style=none] (18) at (-2.5, 0.75) {};
		\node [style=none] (19) at (-1, 0.75) {};
		\node [style=none] (20) at (-0.5, 0.75) {};
		\node [style=none] (21) at (-2, 0.75) {};
		\node [style=none] (22) at (-3, 1.25) {};
		\node [style=none] (23) at (-3, -0.25) {};
		\node [style=none, anchor=mid] (24) at (-2.75, 1.75) {\lang{which}};
		\node [style={small_node}] (25) at (-3, 1.25) {};
		\node [style={small_node}] (26) at (-2.5, 0.75) {};
		\node [style=none] (27) at (-4.5, 0) {$N$};
		\node [style=none] (28) at (1, 0) {$N$};
		\node [style=none] (29) at (-2.75, 0) {$N$};
		\node [style=none] (30) at (-1.5, 0) {$S$};
	\end{pgfonlayer}
	\begin{pgfonlayer}{edgelayer}
		\draw [bend right=90, looseness=0.75] (3.center) to (4.center);
		\draw [bend right=90, looseness=0.75] (5.center) to (6.center);
		\draw [style=swap] (9.center) to (7.center);
		\draw [style=swap] (7.center) to (8.center);
		\draw [style=swap] (8.center) to (9.center);
		\draw [style=swap] (10.center) to (12.center);
		\draw [style=swap] (12.center) to (11.center);
		\draw [style=swap] (11.center) to (10.center);
		\draw [style=swap] (13.center) to (15.center);
		\draw [style=swap] (15.center) to (14.center);
		\draw [style=swap] (14.center) to (13.center);
		\draw [bend right=90, looseness=0.75] (21.center) to (19.center);
		\draw [in=-90, out=-90, looseness=0.75] (18.center) to (20.center);
		\draw (22.center) to (23.center);
		\draw [style=swap, in=180, out=90, looseness=1.00] (16.center) to (22.center);
		\draw [style=swap, in=90, out=0, looseness=1.00] (22.center) to (21.center);
	\end{pgfonlayer}
\end{tikzpicture}

%% file: frob-sub-copy-copy.tikz
\begin{tikzpicture}
	\begin{pgfonlayer}{nodelayer}
		\node [style=none, anchor=mid] (0) at (-3, 1) {\lang{fruit}};
		\node [style=none, anchor=mid] (1) at (-0.5, 1) {\lang{tastes}};
		\node [style=none, anchor=mid] (2) at (2, 1) {\lang{bitter}};
		\node [style=none] (3) at (-3, 0.75) {};
		\node [style=none] (4) at (0, 0.75) {};
		\node [style=none] (5) at (2, 0.75) {};
		\node [style=none] (6) at (3.25, 0.75) {};
		\node [style=none] (7) at (0.75, 0.75) {};
		\node [style=none] (8) at (2, 1.75) {};
		\node [style=none] (9) at (-3, 1.75) {};
		\node [style=none] (10) at (-4.25, 0.75) {};
		\node [style=none] (11) at (-1.75, 0.75) {};
		\node [style=none] (12) at (-0.5, 1.75) {};
		\node [style=none] (13) at (-1.5, 0.75) {};
		\node [style=none] (14) at (0.5, 0.75) {};
		\node [style=none] (15) at (-3, 0.75) {};
		\node [style=none] (16) at (-0.5, 0.25) {};
		\node [style=none] (17) at (-1, 0.75) {};
		\node [style=none] (18) at (-0.5, 0.75) {};
		\node [style=none] (19) at (-1, 0.75) {};
		\node [style=none] (20) at (-2, 0.25) {};
		\node [style=none] (21) at (-2, -0.25) {};
		\node [style={small_node}] (22) at (-2, 0.25) {};
		\node [style={small_node}] (23) at (-0.5, 0.25) {};
	\end{pgfonlayer}
	\begin{pgfonlayer}{edgelayer}
		\draw [bend right=90, looseness=0.75] (4.center) to (5.center);
		\draw [style=swap] (8.center) to (6.center);
		\draw [style=swap] (6.center) to (7.center);
		\draw [style=swap] (7.center) to (8.center);
		\draw [style=swap] (9.center) to (11.center);
		\draw [style=swap] (11.center) to (10.center);
		\draw [style=swap] (10.center) to (9.center);
		\draw [style=swap] (12.center) to (14.center);
		\draw [style=swap] (14.center) to (13.center);
		\draw [style=swap] (13.center) to (12.center);
		\draw [in=-90, out=-90, looseness=0.75] (16.center) to (18.center);
		\draw (20.center) to (21.center);
		\draw [style=swap, in=180, out=-90, looseness=1.00] (15.center) to (20.center);
		\draw [style=swap, in=-90, out=0, looseness=1.00] (20.center) to (19.center);
	\end{pgfonlayer}
\end{tikzpicture}